\documentclass[acmsmall]{acmart}

\usepackage{algpseudocode}
\usepackage{algorithm}
\usepackage{xspace}
\usepackage{listings}
\usepackage{color}
\usepackage{multirow}
\usepackage{dblfloatfix}
\usepackage{lscape} 
\usepackage{xparse}
\usepackage{placeins}
\usepackage{caption} 
\usepackage{subcaption}
\usepackage{enumitem}
\usepackage{array, makecell}
\usepackage{wrapfig}
\usepackage{colortbl}
\usepackage{tabularx} 
\usepackage{booktabs}
\usepackage{courier}
\usepackage{ragged2e}
\usepackage{amsthm}
\usepackage{hyperref}
\usepackage{comment}
\usepackage{pdflscape}
\usepackage{afterpage}
\usepackage{fontawesome}
\usepackage{pifont}
\usepackage{lscape}
\setlist[itemize]{noitemsep, topsep=4pt}
\definecolor{lightgray}{HTML}{f6f6f6}
\definecolor{darkgray}{rgb}{.4,.4,.4}
\definecolor{darkblue}{HTML}{1b4db3}
\definecolor{brickred}{HTML}{b04f4f}
\definecolor{purple}{rgb}{0.65, 0.12, 0.82}
\definecolor{diffadd}{HTML}{288f26}
\definecolor{diffrmbg}{HTML}{ffebe9}
\definecolor{diffaddbg}{HTML}{e6ffeb}
\definecolor{diffremove}{HTML}{de4f54}
\definecolor{carrotorange}{rgb}{0.8, 0.33, 0.0}
\definecolor{highlight}{HTML}{fefbc2}
\lstdefinelanguage{JavaScript}{
  keywords={typeof, new, true, false, catch, function, return, null, catch, switch, var, const, let, extends, if, in, while, do, else, case, break, async, await,of},
  keywordstyle=\color{darkblue}\bfseries,
  ndkeywords={class, export, boolean, throw, implements, import, this, setTimeout},
  ndkeywordstyle=\color{brickred}\bfseries,
  identifierstyle=\color{black},
  sensitive=false,
  comment=[l]{//},
  morecomment=[f][\color{diffadd}\bfseries]{+\ },
  morecomment=[s]{/*}{*/},
  morecomment=[f][\color{diffremove}\bfseries]{- },
  commentstyle=\color{violet}\ttfamily,
  stringstyle=\color{carrotorange}\ttfamily,
  morestring=[b]',
  morestring=[b]"
}

\theoremstyle{definition}

\newcommand{\header}[1]{\par\smallskip\noindent\textbf{#1.}}

\AtBeginDocument{%
	}

\setcopyright{none}

\acmJournal{TOSEM}

\newboolean{showcomments}
\setboolean{showcomments}{true}
\ifthenelse{\boolean{showcomments}}
{
	\definecolor{myyellow}{RGB}{255, 228, 26}
	\definecolor{myblue}{RGB}{50, 50, 220}
	\newcommand{\nb}[2]{
		{\sf
			\fcolorbox{myyellow}{yellow}{\scriptsize\textbf{#1}}%
			$\blacktriangleright$%
			{\color{myblue}\fontsize{7pt}{8pt}\selectfont\textbf{#2}}%
		}%
	}
}
{
	\newcommand{\nb}[2]{}
}

\newcommand{\toolname}{\textsc{Katana}\xspace}
\newcommand{\hoppity}{\textsc{Hoppity}\xspace}

\newcommand{\sequencer}{\textsc{SequenceR}\xspace}
\newcommand{\coconut}{\textsc{CoCoNuT}\xspace}
\newcommand{\code}[1]{{\small\ttfamily\texttt{#1}}}

\makeatletter
\DeclareRobustCommand{\change}{%
  \@bsphack
  \leavevmode
  \color{blue}
  \@esphack
}
\DeclareRobustCommand{\stopchange}{%
  \@bsphack
  \normalcolor
  \@esphack
}
\makeatother

\begin{document}

\title{
\toolname: Dual Slicing-Based Context for Learning Bug Fixes
}


\author{Mifta Sintaha}
\affiliation{%
  \institution{The University of British Columbia}
  \city{Vancouver}
  \country{Canada}}
\email{msintaha@ece.ubc.ca}

\author{Noor Nashid}
\affiliation{%
	\institution{The University of British Columbia}
	\city{Vancouver}
	\country{Canada}}
\email{nashid@ece.ubc.ca}

\author{Ali Mesbah}
\affiliation{%
	\institution{The University of British Columbia}
	\city{Vancouver}
	\country{Canada}}
\email{amesbah@ece.ubc.ca}


\begin{abstract}
	
Contextual information plays a vital role for software developers when understanding and fixing a bug. Consequently, deep learning-based program repair techniques leverage context for bug fixes. However, existing techniques treat context in an arbitrary manner, by extracting code in close proximity of the buggy statement within the enclosing file, class, or method, without any analysis to find actual relations with the bug. To reduce noise, they use a predefined maximum limit on the number of tokens to be used as context. 
We present a program slicing-based approach, in which instead of arbitrarily including code as context, we analyze statements that have a control or data dependency on the buggy statement. We propose a novel concept called \emph{dual slicing}, which leverages the context of both buggy and fixed versions of the code to capture relevant repair ingredients. We present our technique and tool called \toolname, the first to apply slicing-based context for a program repair task. The results show \toolname effectively preserves sufficient information for a model to choose contextual information while reducing noise. We compare against four recent state-of-the-art context-aware program repair techniques. Our results show \toolname fixes between 1.5 to 3.7 times more bugs than existing techniques.
	
\end{abstract}



\maketitle
	
\section{Introduction}

Traditional automated error detection~\cite{google:sa} and program repair~\cite{legouesNFWTSE2012,Scott2019GetafixLT,confix} techniques rely on a set of predefined templates and rules that are limited to specific software bug patterns; adding support for a new type of bug is manual and requires domain-specific knowledge in a given programming language. Instead of hard-coding error detection and repair patterns, we can automatically learn them from code examples of developer made mistakes and repairs, through  deep learning~\cite{DeepFixFC,deepdelta, dlfix,sequencer, coconut}. 

To apply learning-based techniques for software analysis, the source code needs to be vectorized. It is, however, imperative to first delineate what information in the source code  should be included as input for vectorization.  While syntactical representation of source code for vectorization has gained traction in the literature in recent years, semantic information has received less attention. In particular, the notion of \emph{context} pertaining to an erroneous statement in the code has been largely treated in an ad-hoc and arbitrary fashion. For instance, some techniques merely focus on the buggy statement~\cite{pradeloopsla, hata2019learning} and ignore context. Others use the enclosing file~\cite{hoppity,subroutineSummarization}, enclosing class~\cite{sequencer}, enclosing function~\cite{tufano:tosem:19,coconut}, or encapsulating AST subtrees~\cite{dlfix} as context, often with a hard-coded bounding limit on the number of tokens.

Arbitrarily including code as context neither captures the true semantics of the buggy statement nor encompasses essential fix-ingredients. 
For developers, context plays a significant role in understanding a bug and determining a potential fix. This is true in case of machine learning models, where too much information in the input introduces noise that can affect the repair accuracy of the model~\cite{noiseML} and too little information can lead to overfitting or relevant data loss, leading to incorrect generation of patches.
This poses questions pertaining to what is relevant context, how to collect it, what role the repair ingredients should play, and how relevant information related to both the erroneous and fixed code should be represented for deep learning consumption. 

Finding adequate context ingredients is vital to overcoming this limitation.  
In fact, when a developer tries to fix a buggy line of code, they start from the buggy line and examine all the ingredients, such as the variables and function calls used in that line. They then go backward in the code, investigating where such ingredients have been defined, used, and modified throughout the code to understand the error.  Our insight is that such relevant information can help a deep learning approach better reason about bug fixes.   In particular, we propose to adopt backward slicing analysis for extracting  contextual information in the form of code directly related to a buggy statement as well as its repaired ingredients, which we call \emph{dual slicing} in this paper.  Our approach, implemented in a tool called \toolname, extracts  dual slicing-based context from the buggy and fixed files through inter and intraprocedural control and data flow analysis, transforms the slices to AST-based graph representations, and uses a Graph Neural Network (GNN) to train a model. To evaluate our dual slicing approach, we compare against four state-of-the-art  repair techniques, and our results show that \toolname outperforms all four. 

Our work makes the following contributions: 

\begin{itemize}

\item A technique for extracting contextual information to enhance deep learning program repair.  Our program slicing uses control and data flow analysis to find code that is relevant and applies the notion of \emph{dual slicing} to capture context from both the buggy and fixed code. It requires no hard-coded bound on the number of tokens to limit the scope of the context. 

\item  An evaluation of our dual slicing technique and comparison with four state-of-the-art learning-based program repair techniques. We train models on 91,181 pairs of buggy and fixed JavaScript files.  \toolname achieves 42\% ($\frac{4,781}{11,397}$) repair accuracy within top-3 predictions, which is  between 1.5 to 3.7 times higher than existing techniques.

\item A quantitative analysis of the type and amount of information available for a learning model as well as a qualitative report on the type of bugs that \toolname can fix; 891 bugs are exclusively fixed by \toolname. 

\item The implementation of our technique, \toolname, which is available~\cite{katana}; it includes our dataset, model and JavaScript program slicer, compatible with the latest JavaScript ES10.
	
\end{itemize}

\section{Motivation}
Context is essentially the background information related to a particular problem depending on the domain. The importance of context has been discussed in natural language processing tasks~\cite{contextNLP} for establishing semantic similarity between words. For a human developer, context is key in understanding and creating a fix for a buggy program. Therefore, different types of context in both machine learning and rule-based approaches have been applied for fixing a bug~\cite{capgen,dlfix, coconut, sequencer}.   In case of a program repair, the context is typically considered as the surrounding source code relevant to the buggy statement.
Current state-of-the-art learning based program repair approaches employ context in various ways and many of them use sequence-based and tree/graph based source code representation for learning program repair. These approaches use the buggy line with enclosing class \cite{sequencer}, enclosing function \cite{coconut,tufano:tosem:19}, buggy subtree \cite{dlfix}, or the whole file \cite{hoppity} for capturing the context of a buggy line. However, all these approaches have a bounding limit for processing the buggy line. The sequence-based approaches use a maximum token limit (e.g., 1,000 tokens~\cite{sequencer}), within which they extract buggy methods or classes to avoid truncated sequences. Similarly, current graph-based learning techniques use a maximum node limit (e.g., 500 nodes~\cite{hoppity}) for program repair. In practice, we cannot expect real world bugs to always be enclosed within small/medium sized methods or be limited to an arbitrary number of tokens or AST nodes. Therefore, this kind of quantitative constraint can potentially cause some important information pertaining to the bug fixes to be truncated and the overall context may even lose semantic meaning. In addition, increasing the limit on the number of tokens or nodes may add noise to the model.

\begin{lstlisting}[firstnumber=1]
'use strict';
import ENV from 'pass-ember/config/environment';
\end{lstlisting}	  
\vspace{-\baselineskip}
\begin{lstlisting}[firstnumber=3, backgroundcolor=\color{highlight}]
import { get } from '@ember/object';
import CheckSessionRoute from '../../check-session-route';
function service(user) {
	return {
		...user,
		userToken: get('currentUser.accessKey'),
		userSecret:  get('currentUser.userSecret'),
	};
}
const user = get('currentUser.user');
export default CheckSessionRoute.extend({
-   currentUser: service(),
+   currentUser: service(user),
\end{lstlisting}
\vspace{-\baselineskip}
\begin{lstlisting}[firstnumber=15]
	model() {
		...
	}
\end{lstlisting}
\vspace{-\baselineskip}
\begin{lstlisting}[firstnumber=30]
	...
\end{lstlisting}
\vspace{-\baselineskip}
\begin{lstlisting}[firstnumber=66, backgroundcolor=\color{highlight}, label={lst:mot1}, caption={Sliced program with missing function argument},captionpos=b]
}); 
\end{lstlisting}

We will use Listing \ref{lst:mot1}, which shows a real bug and its fix for a JavaScript project on GitHub, as a running example. This type of bug is common in many languages and falls under \emph{Same function more args} pattern as categorized by Karampatsis and  Sutton~\cite{minesstubs}. Let us consider how a developer would go about fixing the bug here. After localizing the faulty line of code, they would try to determine the root cause of the bug. To do so, they would analyze how the variables or functions in the faulty line have been affected in the previous lines. The developer notices that the property \code{currentUser} is invoking a function call to \code{service}. After examining the function declaration, they notice that \code{service} expects a required \code{user} argument and that the buggy line is not passing any arguments. The developer identifies the cause of the fault and generates a fix by passing the missing argument, \code{user} declared in line 12, to the function call. 

Having too large of a context, especially if the method/class enclosing the buggy line is huge or has many dependencies can cause significant overhead~\cite{developersBugs}. 
By limiting the scope to only lines relevant to the fault, the developer can avoid information overload during the process of debugging and repairing the program.

In Listing \ref{lst:mot1}, the highlighted code indicates the portion relevant for the developer to understand and fix the bug in line 14.
The original buggy file in this example contained 66 lines of code and by focusing only on the relevant lines, the attention decreases to 13 lines. In practice, these kinds of bugs are pertinent in much more complex settings containing many lines and functions with more arguments. Furthermore, if we were to follow the approach of some of the current deep learning models, and extract only the enclosed method or enclosed class of the buggy line, then we would lose meaningful context outside the enclosing scope like the \code{user} variable and \code{service} function that were necessary for the developer in generating the correct fix. 

Our insight is that for context, instead of constraining the attention to the enclosing entities or arbitrary limitations on the number of tokens/nodes, we can constrain the code to those portions that are relevant to the bug and fix through the notion of program slicing. 

\section{Approach}

\begin{figure*}[h]
	\centering
	\includegraphics[width=\textwidth]{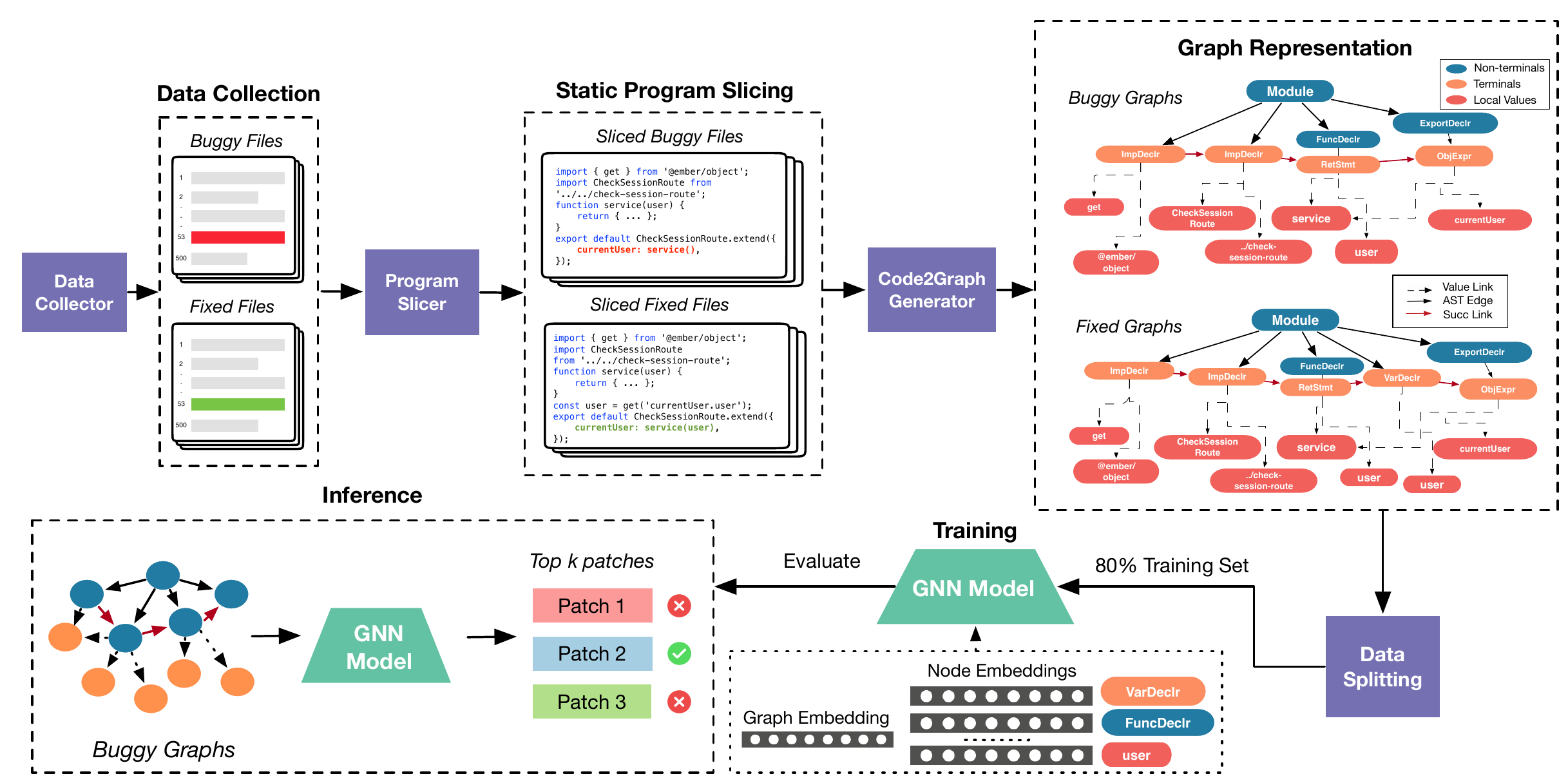}
	\caption{Overview of our approach.}
	\label{fig:overview}
\end{figure*}

We hypothesize that slicing-based context can be helpful to a machine learning model that learns patterns of bug fixes. Our technique, called \toolname, is based on static program slicing to obtain relevant context with respect to the bug and its fix. This obtained slice-based context is then represented as an AST-based graph augmented with control and data flow edges. This graph is then fed into a Graph Neural Network (GNN) for training. 
Figure \ref{fig:overview} depicts the  overview of our approach. Our overall approach consists of five main steps, namely, data collection, program slicing, graph representation, training, and inference. Next, we describe each step.

\subsection{Data Collection}
To collect data, we chose JavaScript as the target language. According to recent StackOverflow Developer Surveys~\cite{stackOSurvey}, JavaScript stands as the most commonly used language in the world as reported by 68.62\% of professional developers. JavaScript was initially limited to client-side code running in the web browser but has gained popularity in backend development through the emergence of the Node.js\footnote{https://nodejs.org} framework. 
Unlike statically typed languages such as Java or C\# where many errors can be discovered at compile time, JavaScript tends to "swallow'' errors and silently continue execution. Therefore, having a good program repair tool for a dynamically typed language such as JavaScript can help developers with a seamless coding experience.

We collected pairs of buggy and fixed JavaScript files by crawling commits of open-source JavaScript repositories hosted on GitHub. We narrowed down the search by focusing on  commits containing the keywords "bug", "fix"  and "resolve". We collected data from top 1,135 JavaScript GitHub projects based on the number of stars. We exclude duplicates from the test dataset, as well as any data points that appear in both the test and training datasets~\cite{allamanis2019adverse}. Additionally, we use heuristics to filter out commits that may be feature additions or refactorings based on the number of AST differences. To this end, we consider a difference of one AST node to be a bug, similar to existing work~\cite{hoppity}. A datapoint is considered to have one node difference if only a single AST node has been added, removed or modified in the buggy tree to obtain the fixed tree. This includes any internal node or leaf node e.g. operators, variables or values. Listing~\ref{lst:onenode} demonstrates the examples of such cases. We extract these datapoints by 1) generating the AST of buggy and fixed files of each datapoint using SHIFT AST\footnote{https://shift-ast.org} parser and, 2) selecting only those datapoints where the diff between the two ASTs is equal to one. We exclude minified JavaScript files since our program analysis tool cannot handle those. We also prune any unparseable JavaScript datapoints using the SHIFT AST parser. If a JavaScript file has incomplete closures such as missing braces or parenthesis, we use the parser to check if an AST can be generated or not. We collected a total of 113,975 datapoints, i.e., pairs of buggy and fixed JavaScript files having one AST node difference between the pairs.

\begin{lstlisting}[label={lst:onenode},,caption={Example of one node difference}]
-  sum(a, b);
+  sum(a, b, c);

-  return a - b;
+  return a + b;
\end{lstlisting}

\subsection{Static Program Slicing}
\label{sec:statprogslicing}
We propose to use static analysis to examine the buggy line and all the ingredients, such as the variables and function calls used in that line, and backward program slicing~ \cite{weiser}, to determine  where such ingredients have been defined, used, and modified throughout the code. We use this backward slice as context in our work, which is a subset of all statements with data or control dependencies with respect to the slicing criterion $(p,V)$, which is the set of variables $V$ at the buggy program point $p$. 

{\small
	\renewcommand{\algorithmicrequire}{\textbf{Input:}}
	\renewcommand{\algorithmicensure}{\textbf{Output:}}
	\algnewcommand\algorithmicforeach{\textbf{foreach}}
	\algdef{S}[FOR]{ForEach}[1]{\algorithmicforeach\ #1\ \algorithmicdo}
	\begin{algorithm} \caption{Extracting Backward Sliced Context}\label{alg:backwardslice}
		\begin{algorithmic}[1]
		\Require $N$: A given line number; $file$: A given file
		\Ensure List: String: List of statements
		\State $entitySet \gets getEntities(N, file)$
		\State $contextLines \gets \{N\}$
		\ForEach {$entity \in entitySet $}
		\State \Call{BackwardSlice}{$entity$, $contextLines$}
		\EndFor
		\State \Return $getStatements(contextLines)$
		\space
		\Function{BackwardSlice}{$ent$, $contextLines$}
		\ForEach {$ref \in ent.refs() $}
		\If{$ref.ent()$ is control-dependent or data-dependent on $ent$}
		\State $N \gets ref.line()$
		\State $contextLines \gets contextLines \cup N$
		\State $entitySet \gets getEntities(N, file)$
		\ForEach {$entity \in entitySet $}
		\State \Call{BackwardSlice}{$entity$, $contextLines$}
		\EndFor
		\EndIf
		\EndFor
		\EndFunction
		\end{algorithmic}
	\end{algorithm}
}

There are two types of program slicing analysis, namely, inter-procedural and intra-procedural analysis. Intra-procedural analysis is performed within the scope of the function and inter-procedural analysis is performed within the whole program. We opted for performing both inter-procedural and intra-procedural analysis but limiting the slicing scope within the JavaScript file. The reason for such a choice is that only 12\% (13,964) of the 113,975 datapoints have the buggy line enclosed within a function and 3\% (22,940)  have the buggy line as part of a class. In 85\% (97,071), the buggy line is part of the global scope. 
In JavaScript, the global code can access all the constructs within a function, and the global code supports stepwise execution, just like functions do. This is unlike languages such as Java where the \code{main} method is required as the entry point for code execution. For this reason, we go beyond intra-procedural program analysis to capture a broad spectrum of datapoints. We kept our scope of analysis limited to the file itself, similar to other approaches~\cite{coconut,hoppity,sequencer,tufano:tosem:19}. However, unlike these approaches, we do not apply any adhoc bounding limits and instead use backward slicing to capture repair ingredients systematically. This is because a limit based on an arbitrary number of tokens can potentially miss relevant context, particularly in cases where proximity to the contextual statements are above the bounding limit.

We analyze each datapoint in our dataset to collect the necessary information to conduct the slicing analysis. In order to determine the slicing criterion, we first perform a diff analysis between the buggy and fixed lines to obtain the buggy line number. We then feed this line number to our slicing framework and extract the variables and function calls used in that line. For extracting the contextual statements through static backward slicing, we incorporate both control flow and data flow analysis and identify the sliced statements where applicable. 

Algorithm~\autoref{alg:backwardslice} illustrates a high-level description of the steps used in extracting the context from a given line number, $N$ and $file$. Specifically, Algorithm~\autoref{alg:backwardslice} extracts the entities (i.e., variables, functions, objects) from the given buggy or fixed line in a file. Algorithm~\autoref{alg:backwardslice} then iterates through all the entities to obtain all the $contextLines$ using backward slicing analysis. $contextLines$ is a set of line numbers of the sliced context. In the $BackwardSlice$ function, Algorithm~\autoref{alg:backwardslice} iterates through all the references in a given entity; if the referenced entity is control or data-dependent, then the corresponding line is added to $contextLines$; this process continues recursively to obtain the backward slice of the referenced entities. Finally, the algorithm constructs the source code statements from the $contextLines$ to return the output. Here, the $getStatements$ function ensures that all the closures of a given context line are extracted correctly, as it is a challenging process in a dynamic language such as JavaScript. For example, unlike Java, a semicolon is not required in JavaScript for declaration or expression statements; in such cases, we consider the new line as the end of a statement. The buggy line is within a loop or conditional statement in some cases. In such buggy statements, we determine the closure through control flow analysis to avoid an incomplete context with backward slicing.

\begin{lstlisting}[label={lst:buggySliced}, caption={Sliced Buggy File},captionpos=b]
import { get } from '@ember/object';
import CheckSessionRoute from '../../check-session-route';
function service(user) {
	return {
		...user,
		userToken: get('currentUser.accessKey'),
		userSecret:  get('currentUser.userSecret'),
	};
}
export default CheckSessionRoute.extend({
-	 currentUser: service(),
});  
\end{lstlisting}

Our extracted datapoints of buggy and fixed files contain the whole JavaScript code with only the bug, and the patch that contains the difference at the same buggy line. We propose two types of slicing mechanisms, namely single slicing and dual slicing. For each, we generate a separate set of datasets and conduct experiments to measure their effectiveness in repairing programs. These two types of slicing mechanisms are described below.

\header{Single Slicing}
Single slicing analysis is used for extracting context with respect to the buggy line. We start off by first taking the buggy line as the slicing criterion to capture the backward slice statements as context. This generates a buggy file with only the backward sliced statements as context. We then attach this context of the buggy file to the correct line and generate the fixed version of the buggy file. We call this \emph{single slicing} because backward slicing has been applied only on the buggy file and the context of the buggy file is simply transferred to the fixed file. Listing \ref{lst:buggySliced} and \ref{lst:singleSliced} demonstrate the snippets of single sliced pairs of buggy and fixed files from the same example shown in Listing \ref{lst:mot1}. As the slicing criterion was the buggy line, the fixed file contains the same sliced context of the buggy file and contains no information about the variable \code{user} that has been passed as a fix.

\begin{lstlisting}[label={lst:singleSliced}, caption={Single Sliced Fixed File},captionpos=b]
import { get } from '@ember/object';
import CheckSessionRoute from '../../check-session-route';
function service(user) {
	return {
		...user,
		userToken: get('currentUser.accessKey'),
		userSecret:  get('currentUser.userSecret'),
	};
}
export default CheckSessionRoute.extend({
+   currentUser: service(user),
}); 
\end{lstlisting}

\begin{lstlisting}[label={lst:dualSliced}, caption={Dual Sliced Fixed File},captionpos=b]
import { get } from '@ember/object';
import CheckSessionRoute from '../../check-session-route';
function service(user) {
	return {
		...user,
		userToken: get('currentUser.accessKey'),
		userSecret:  get('currentUser.userSecret'),
	};
}
const user = get('currentUser.user'); // NEW CONTEXT
export default CheckSessionRoute.extend({
+   currentUser: service(user),
});
\end{lstlisting}

\header{Dual Slicing} 
Our intuition is that, certain repair ingredients are available in the fix context, which can  improve the overall learning process. For instance, if the fix introduces a new identifier that was not present in the buggy line, then the corresponding context will be different in the fixed version of the file. This pattern is known as \emph{Change Identifier Used}, and is commonly observed in many programming languages~\cite{minesstubs, PySStuBs}. Learning from the context of both the buggy and fixed code can provide additional semantic information to the model.  

Therefore, we hypothesize that extracting slices from both the buggy and fixed files can be more effective in conveying the contextual information to the training model. We call this notion of collecting buggy and repair contextual ingredients \emph{dual slicing}. Dual slicing analysis aims for extracting context separately from both buggy and fixed files. By using the slicing criterion from the buggy line and fixed line, we extract the sliced statements from both these lines individually to generate the sliced buggy and fixed files. Listing \ref{lst:buggySliced} and \ref{lst:dualSliced} demonstrate the dual sliced pairs of the buggy and fixed files from the motivating example of \autoref{lst:mot1}. As the slicing criteria are both the buggy and fixed lines in this approach, the fixed file contains an additional line containing the variable declaration of \code{user} in line 10 which is necessary for generating the patch as the function \code{service} expects this param.

\subsection{Graph Representation}
Once we have prepared the single sliced and dual sliced datapoints, we move to the learning phase of the approach. To represent source code, we opt for graph representation. Graph representations of source code have gained popularity recently due to their ability to represent semantic information~\cite{allamanis-represent-programs-with-graphs-2018, hoppity, devign-nips-vulnerability-detection-gnn,het-graph-icpc-2022}. Typically, the program's abstract syntax tree (AST) is used as the backbone of a program graph which consists of syntax non-terminals and terminals of a programming language's grammar. Graphs are able to leverage both the syntactic and semantic relations between the nodes via different types of edges.  They are also able to consider long-range dependencies through edges, between same variables or functions even if they are placed at distant locations ~\cite{allamanis-represent-programs-with-graphs-2018}. This is unlike sequence based approaches which can sometimes lose meaningful contextual information due to a maximum token limit.

For each datapoint, we create graphs for the buggy and fixed sliced files separately. Following previous work~\cite{hoppity}, we extract the AST of each sliced files and convert it to a graph with the addition of \code{SuccToken} edges to connect leaf nodes and \code{ValueLink} edges to connect additional value nodes. Figure \ref{fig:overview} illustrates the buggy and fixed graphs for the slices of our example (\autoref{lst:mot1}). After representing the source code as a graph, the resulting graphs are mapped into a vector representation using a Graph Neural Network (GNN). More specifically, given a graph $g = (V, E)$ with a set of nodes $V$ and edges $E$, we determine the $d$-dimensional representation of graph $g$ and individual node representations $\upsilon \in V$ using a function $f(g) \mapsto (\mathbb{R}^{d},\mathbb{R}^{|V| \times d})$. The node embedding is a vector, $\vec{v} = h_v^{(L)}$ where $L$ denotes the total number of propagations in the GNN via message passing \cite{gnn}. Message passing is used in GNN models in which vector messages are exchanged between nodes in the graph and updated through an aggregation function. The graph representation $\vec{g}$ is the aggregation of node embeddings $h^l_v,\forall l \in 0, 1,...,L$. To aggregate $h^l_v$ for each $l$, max pooling is used, which takes the average of the $L + 1$ vectors to obtain $\vec{g}$. 

\subsection{Training}
\label{subsec:training}
In this step, we split the graphs generated from the previous step into training, testing and validation using random sampling. We fold the dataset into 80\% training set, 10\% testing set, and 10\% validation set. Table \ref{table:data_split} provides the number of datapoints in the training, validation, and test datasets in our study. To train a model, we adopt a GNN architecture that maps the graph representation into a fixed dimensional vector space. The program repair is essentially a series of graph transformation operations, from the buggy graph to the fixed graph, containing operations such as adding or deleting a node, replacing a node value or node type. During the graph creation step, the buggy graph ($g_{bug}$), fixed graph ($g_{fix}$) and the corresponding sequence of graph edits to create the $g_{fix}$, i.e., the graph diff is generated for each datapoint. In other words, the model creates $g_{fix}$ by applying graph edit operations to the $g_{bug}$, on the buggy line. This graph diff comprises a sequence of AST modifications which serves as a fine-grained supervision mechanism for graph transformation. A modification contains the type of graph edit operation, node location in the graph and value to be added/modified (in case of deletion, no value is provided). To summarize, given a dataset $ \mathnormal{D} = \{(g^{(i)}_{bug}, g^{(i)}_{fix})\}^{|\mathnormal{D}|}_{i=1}$ containing pairs of buggy and fixed sliced code, the model is trained with the learning objective, $max_\theta\mathbb{E}_{(g_{bug},g_{fix})\sim D}p(g_{bug}| g_{fix};\theta)$ to maximize the likelihood of fixes~\cite{hoppity}. The Maximum Likelihood Estimation (MLE) objective is calculated as the sum of cross-entropy loss at each step of graph edits. 

We train two separate models for each types of slicing. In single and dual slicing, the input and output to the model varies in terms of the contextual information. For single slicing, the input to the GNN model is the buggy graph, $g_{bug}$ which contains the sliced context with respect to the buggy line only. The output of the model is a graph edit operation applied to the buggy node of $g_{bug}$, resulting in $g_{fix}$. On the other hand, for dual slicing, the model input is the buggy graph, $g_{bug}$ which contains the sliced context with respect to the buggy line along with the fixed line. The output remains same as before i.e., a graph edit operation applied to the buggy node in $g_{bug}$. In short, the model trained with dual sliced context has access to extra contextual information pertaining to both the bug and fix. Note that, we do not differentiate the buggy or fixed graph input to the model based on any particular bug type, and feed the data as is.

\begin{table}[h]
	\small
	\centering
	\caption{Dataset split for training, testing, and validation. The numbers for graph edit type are also indicated.}
	{
		\begin{tabular}{lrrrrr}
			\toprule
			& \code{ADD\_NODE} & \code{DEL\_NODE} & \code{REP\_TYPE} & \code{REP\_VAL} & \textbf{Total} \\
			\cmidrule[0.4pt]{1-6}
			Train & 1,949 & 2,632 & 1,061 & 85,539 & \textbf{91,181}\\
			Validate & 237 & 343 &  198 & 10,619 & \textbf{11,397} \\
			Test & 243 & 370 & 209 & 10,575 & \textbf{11,397} \\
			\bottomrule
		\end{tabular}
	}
	\label{table:data_split}
\end{table}

\header{Hyperparameter Tuning} 
Hyperparameters are important since they influence a model's overall performance. We use a batch size of 10 and tune the hyperparameters of the model for the sliced datapoints. In order to find the optimum set of hyperparameters, we perform a manual search for discrete values of the number of GNN layers, learning rates, and dropout. To this end, we picked 2, 3 and 4 layers for the GNN model, learning rates of 0.1, 0.01, 0.001 and dropout values of 0.0, 0.1, and 0.2. We trained the model on 5 epochs with a batch size of 10 for each combination of hyperparameters. Overall, we trained 27 models during the hyperparameter search and found the best performance using 4 GNN layers, a learning rate of 0.001 and a dropout of 0.1.

\subsection{Inference}
After the training step, we evaluate each model using the test dataset. Given a buggy file during inference, we assume that the buggy line has already been identified in a fault localization step, similar to the current state-of-the-art learning-based repair techniques~\cite{sequencer,dlfix,coconut}. Locating the buggy line is a necessary pre-processing step to extract the backward slice as context. The model localizes the bug within the buggy graph and generates the corresponding patch. The input and output to the models during evaluation are as below: 

\begin{itemize}
\item \emph{Training}: As described in Section~\ref{subsec:training}, we train two separate models with the single slicing and dual slicing datasets.
\item \emph{Inference}: Input to the models remains the same during inference. For evaluating the trained single and dual sliced models, input to the model is a graph, $g_{bug}$ which is generated from the single-sliced buggy JavaScript file containing the buggy line. 
\item \emph{Output}: For both cases, the model output is a series of graph edits applied to $g_{bug}$ resulting into a fixed graph, $g_{fix}$. The model tries to generate $T$ steps of transformations, where $T$ denotes the number of graph edits. We set the value of $T$ to 1 as our dataset contains bugs with one AST node difference.
\end{itemize}

For generating the fix for a $g_{bug}$, both single and dual slice models maintain a pool of 492 node types to predict the type; there are 5,001 values for the single slice model and 5,002 for the dual slice model based on the frequency of value tokens, in a global value dictionary to predict node values; these types and value tokens are extracted from our training datasets. The extra vocabulary in the dual sliced model is \code{header-wrap}. It is the class name of a declared style used within React components. In this way, the vocabulary contains value tokens from both the buggy and fixed graphs, constructed from the buggy and fixed lines along with their sliced context. When generating patches to replace a node value, 
if a vocabulary is not in the global value dictionary, the patch is replaced with \code{UNKNOWN}. In the inference step, the model generates the top $k$ patches for a buggy program as a series of graph edits depending on the beam size. During inference step for both single and dual slicing, the input to the model is a single sliced buggy graph $g_{bug}$.
We consider an inferred patch as correct if the type, location and value of the fix has been accurately identified. This is achieved by matching the actual graph diff from the graph generation step with the inferred graph diff. For example, in case of \autoref{lst:mot1}, the patch prediction is considered correct because the value ("user"), operation type (\code{ADD\_NODE}) and AST node location (node index 44 in the AST) was correctly identified during inference.

\subsection{Implementation}
We implemented our technique in a tool called \toolname~\cite{katana}. Since we were not able to find any existing program slicing tool that supports the latest JavaScript ES10 features, as required for analyzing our dataset, we implemented our own JavaScript slicer for \toolname. Following previous work~\cite{capgen}, we build on top of the Understand~\cite{und} program analysis framework to analyze control flow and data flow dependencies in JavaScript code with respect to our slicing criterion. The slicer in \toolname takes as input a buggy JavaScript file, and the slicing criterion, which is the buggy line and the entities (variables, objects or functions) used in that line. It produces as output a sliced JavaScript file, which is used as context for learning repairs. The deep learning model in \toolname is based on the GNN architecture of \hoppity~\cite{hoppity}. 

\section{Evaluation}
\label{sec:evaluation}

To assess the effectiveness of \toolname we address the following research questions: 

\newlist{researchquestions}{enumerate}{1}
\setlist[researchquestions]{label*=\textbf{RQ\arabic*}}

\begin{researchquestions}
	
	\item How does dual slicing compare to single slicing for learning repairs?

	\item How does \toolname compare to state-of-the-art learning-based repair techniques?
	
	\item What is the effect of slicing on the information obtained as context? 
	 
\end{researchquestions} 

We discuss the experiments that we designed to answer each of the research questions in the following subsections and outline the results.

\subsection{Single versus dual slicing as context (RQ1)}
\label{sec:rq1}
We compare the effectiveness of single slicing where we extract context from the buggy file with dual slicing where the context is extracted from both the buggy and fixed files. 

Using our initial 91,181 (training) datapoints, we generated two separate datasets using our single and dual slicing techniques. The processing time to generate slicing is negligible. On average, single slicing and dual-slicing analysis takes around 330 and 490 milliseconds for a single datapoint, respectively. With the tuned set of hyperparameters, we trained two models separately with the generated single sliced and dual sliced datasets. It took 1.7 hours to train each model for 50 epochs. The average inference time for a datapoint is around 2.54 seconds for each trained model. All our experimental runs are performed on an Ubuntu Linux 18.04.2 LTS server with 122 GB RAM, 16 vCPUs and 2,000 GB SSD. 

\begin{table}
	\small
	\centering
	\setlength{\aboverulesep}{0pt}
	\setlength{\belowrulesep}{0pt}
	\caption{Accuracy of single slicing vs. dual slicing}
	{
		\begin{tabular}{l | rrr}
			\toprule
			\multirow{2}{*}{\textbf{Approach}} 	& \multicolumn{3}{c}{\textbf{Accuracy}} \\
			\cmidrule[0.3pt]{2-4}
			& \textbf{Top-1}    & \textbf{Top-3 }  & \textbf{Top-5}   \\
			\cmidrule[0.3pt]{1-4}
			Single Slicing	& 20.72\% &  35.01\% & 42.90\%  \\
			Dual Slicing & 28.31\% & 41.95\%  & 46.96\%  \\
			\bottomrule
		\end{tabular}
	}
	\label{table:rq1}
\end{table}

We use the same test dataset to evaluate the accuracy of the two models. As the model produces a series of graph edits to generate the patch, we consider an output to be correct if the predicted location of the AST node, the type of operation (i.e., replace type or value, add node, delete node) and the value of the node matches the expected node location, operation type, and value of the patch in the test dataset, for the entire sequence of graph edits. 

Table \ref{table:rq1} shows the results for single and dual slicing in \toolname. 
Single sliced context yielded accuracies of 20.72\%, 35.01\%, and 42.90\% with beam sizes of one, three, and five, respectively. 
Dual sliced context yielded accuracies of 28.31\%, 41.95\%, and 46.96\%, respectively.  Dual slicing achieved higher accuracies for all beam sizes, with 37, 20, and 9 percent increases for the top-1, top-3, and top-5 suggestions, respectively, with respect to single slicing. 

For instance, if we focus on top-3 suggestions, out of the 11,397 bugs in the test dataset, dual slicing fixed 4,781 and single slicing fixed 3,990. Out of these, 3,984 were fixed by both single and dual slicing, however, the dual slicing approach was able to fix 797 more bugs than single slicing.

\vspace{0.5em}
\begin{lstlisting}[backgroundcolor=\color{white},frame=1,label={lst:cbpatch},frame=single,
	framesep=5pt,backgroundcolor=\color{lightgray},
	rulecolor=\color{lightgray}, numbers=none,label={lst:dualfix},caption={Example of \emph{Change Identifier Used} bug pattern fixed by dual slicing}]
-   cart.push(object);
+   cart.push(item);
\end{lstlisting}
\vspace{0.5em}

Listing \ref{lst:dualfix} shows an example of a bug that was correctly fixed by dual slicing but was not repaired by single slicing. Through sampling, we verified that \emph{Change Identifier Used} is one of the recurring patterns among the correct patches generated by dual slicing. We attribute this pattern to the availability of repair ingredients, made accessible through dual slicing as context used during training. Therefore, since the dual sliced context outperforms the single sliced context, we select it as our default approach in \toolname going forward.

\subsection{Comparison with state-of-the-art (RQ2)}

We compare \toolname with the four recent deep learning-based program repair techniques. We selected these state-of-the-art models based on the criteria that, (a) the technique supports JavaScript, (b) the artefact is available, and (c) if the technique does not support JavaScript, at the very least, it can be reimplemented and retrained for JavaScript. To this end, we compared \toolname with four state-of-the-art models. Among them, \hoppity~\cite{hoppity} and \coconut~\cite{coconut} both support JavaScript and their artefacts are available for retraining a model from scratch using our dataset. On the other hand, Tufano et al.~\cite{tufano:tosem:19} and \sequencer~\cite{sequencer} do not support JavaScript and were only trained on Java. However, these techniques are sequence based and do not rely on program analysis or language-specific features. Therefore, adding support for JavaScript was possible within their framework. In addition, we considered comparing \toolname to    CURE~\cite{cure-program-repair-icse-2021}, Recoder~\cite{recoder-fse-2021}, DLFix~\cite{dlfix}, DEAR~\cite{dear-icse-2022}, and RewardRepair~\cite{rewardrepair-icse-2022} using our dataset. Among these techniques, CURE, Recoder and RewardRepair rely on language-specific features and currently only support Java. While we investigated how to train DLFix and DEAR on our dataset, their artifacts were not available at the time of writing this manuscript (the authors confirmed in response to our emails).

A brief outline of these state-of-the-art deep learning-based program repair techniques is described below:

\begin{enumerate}	
	\item \textbf{Tufano et al. \cite{tufano:tosem:19}} employs code abstraction on an RNN-based NMT model with attention mechanism.
	\item \textbf{\sequencer \cite{sequencer}} uses an RNN-based NMT model equipped with copy mechanism.
	\item \textbf{\hoppity \cite{hoppity}} proposes a GNN model to predict the location of bug and generate a fix through a sequence of graph edits.
	\item \textbf{\coconut \cite{coconut}} leverages a CNN-based NMT model with two separate encoders for buggy line and context.
\end{enumerate}

\subsubsection{Characterizing context}
To characterize how current techniques treat context, we reason about five different aspects of context:

\begin{itemize}
	\item \textbf{Analysis}: Contextual information can be extracted from the source code in various ways. The analysis can be as simple as naively extracting tokens from the buggy statement's surroundings, to more complex program analysis techniques such as data flow, control flow, or slicing. 
	
	\item \textbf{Representation}: Once extracted, context can be represented in different ways, e.g., as linear sequence of tokens, or nodes in a graph.
	
	\item \textbf{Scope}: Different levels of granularity can be considered for the scope of context, for example, by focusing on the entire program, enclosing file, class, or method. 
	
	\item  \textbf{Proximity}: Context can be extracted with respect to the location of the bug/fix, e.g., surrounding, before, or after the buggy line.
	
	\item \textbf{Limit}: Context size can be limited by the amount of information it contains, e.g., number of lines of code, tokens, or AST nodes. 
	
\end{itemize}

\begin{table}[ht]
	\scriptsize
	\centering
	\setlength{\aboverulesep}{0pt}
	\setlength{\belowrulesep}{0pt}
	\caption{Accuracy comparison of \toolname with other learning-based program repair techniques.}
	\resizebox{\textwidth}{!}		
	{
		\begin{tabular}{l | lllll | rrr}
			\toprule
			\multirow{2}{*}{\textbf{Approach}} 	& \multicolumn{5}{c|}{\textbf{Context}}	& \multicolumn{3}{c}{\textbf{Accuracy}} \\
			\cmidrule[0.3pt]{2-9}
			& \textbf{Analysis} & \textbf{Representation} & \textbf{Scope} & \textbf{Proximity} & \textbf{Limit} 						& \textbf{Top-1}    & \textbf{Top-3 }  & \textbf{Top-5}   \\
			\cmidrule[0.3pt]{1-9}
			Tufano et al.~\cite{tufano:tosem:19}		 		& Naive 					& Sequences of tokens		& 	Enclosing method 					& Before/After				 	&	100 tokens 			& 4.90\% 					& 11.33\% & 15.84\%  \\
			\sequencer~\cite{sequencer} 				& Naive						& Sequences of tokens		&  Enclosing class  & Before/After					&  1000 tokens 			&7.99\% 					& 13.84\% & 18.08\%  \\
			\hoppity~\cite{hoppity} 					& Naive						& AST-based graph						&  Enclosing file  						& Before/After					&  500 nodes	 		& 5.05\% & 14.40\% & 19.62\% \\
			\coconut~~\cite{coconut}  					& Naive						& Sequences of tokens		&	Enclosing method 					&	Before/After					& 1,022 tokens 			& 24.67\% & 27.90\% & 28.97\% \\			
			\textbf{\toolname} 	& \textbf{Dual Slice} 			& \textbf{AST-based graph} 					&  \textbf{Enclosing file}  						& 	\textbf{Before}							&	\textbf{N/A}					&	\textbf{28.31\%} &\textbf{41.95\%}  & \textbf{46.96\%}  \\
			\bottomrule
		\end{tabular}
	}
	\label{table:rq3}
\end{table}

Table~\ref{table:rq3} compares how state-of-the-art techniques treat context using these five aspects. 

\subsubsection{Setup} 
We trained a separate model using the deep learning framework of each techniques following the dataset split of Table~\ref{table:data_split} with our unsliced dataset as follows:

Tufano et al. \cite{tufano:tosem:19} represent source code as a sequence of tokens and use the method enclosing the bug as the scope of context, which is limited by a maximum of 100 tokens. There is no program analysis for extracting context. However, abstraction is applied to code to limit the vocabulary size. We applied the same abstraction technique to our JavaScript dataset. We tokenized the buggy and fixed JavaScript files and tokenized it using js-tokens\footnote{https://www.npmjs.com/package/js-tokens} to discern whether a given token is an identifier, method, or a literal. Following the same approach as ~\cite{tufano:tosem:19}, we maintain a dictionary of frequently occurring tokens and replace the remaining tokens with abstraction depending on the token type (e.g.,  METHOD\_1, VARIABLE\_1). Additional challenges with JavaScript code included the presence of JSX identifiers and literals. We assigned new abstractions for these types of tokens. The final vocabulary size became 815, including abstractions, JavaScript keywords, and frequent tokens. 

\sequencer represents source code as a sequence of tokens. However, they use the enclosing class as context scope, limited by  1,000 tokens; it delineates the buggy line within <START\_BUG> and <END\_BUG> tokens for differentiation within the context. \sequencer takes more input tokens from lines that appear before the buggy line. It extracts 2/3rd more tokens from proceeding statements before the buggy line than the subsequent statements. We applied the same technique to our JavaScript code corpus, and if the buggy line was not encapsulated in a class, we extracted tokens from the surrounding lines in the JavaScript file. The vocabulary size is limited to 1,000 tokens.

\hoppity represents context as an AST-based graph and does not apply any program analysis. Its context scope is the enclosing file of the buggy line, limited to 500 AST nodes. The vocabulary size on our dataset is 5,003 tokens.

\coconut leverages the enclosed method of the buggy line as the scope of context and represents the source code as a sequence of tokens limited by 1,022 tokens. Similar to other techniques, context is taken without any program analysis. They employ abstraction and use two separate encoders to feed the buggy line and context. We used pre-processing scripts from their artefact to tokenize source code, and the resulting vocabulary size was 63,499.

We use the same test dataset of 11,397 JavaScript bugs for evaluating all the models. The best reported hyperparameter settings and epochs were used to train each of these models. To keep the accuracy comparison equitable across all the models, we use beam sizes of one, three, and five. 

\subsubsection{Results} 
Table  \ref{table:rq3} reports the obtained accuracies for beam sizes of one, three, and five in the columns \emph{Top-1}, \emph{Top-3} and \emph{Top-5}. The lowest \emph{Top-1} accuracy was yielded by Tufano et al.~\cite{tufano:tosem:19}. Despite using a high level of abstraction and a small vocabulary size, limited contextual information has not helped their model. As more contextual information was fed into the model, accuracy improved gradually. \hoppity demonstrated better accuracy for \emph{Top-3} and \emph{Top-5} predictions. Among existing techniques, \coconut yielded the highest accuracy. However, \toolname outperforms all existing techniques for all the beam sizes, with accuracies of 28.31\% , 41.95\%, and 46.96\%  for the beam width 1, 3, and 5, respectively.

Considering the top-3, \toolname is 270.26\%, 203.11\%, 191.32\%, 50.36\% more accurate in fixing buggy programs than Tufano et al.~\cite{tufano:tosem:19}, \sequencer, \hoppity, and \coconut, respectively. By learning from the context of the relevant statements in the buggy and fixed file, the dual slicing-based context in \toolname can accurately repair buggy programs by a significant margin.

\begin{figure}
	\centering
	\includegraphics[width=244pt]{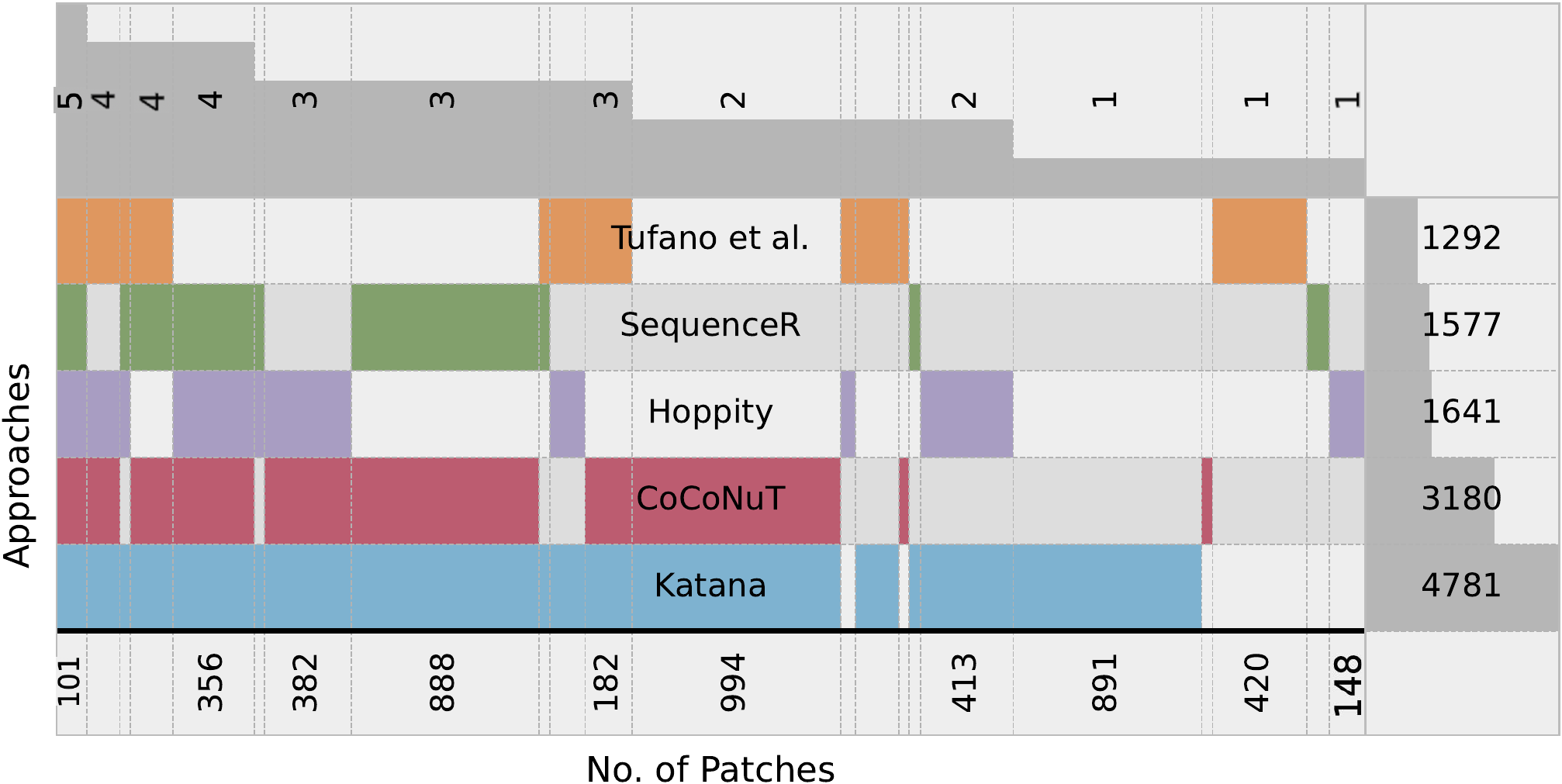}
	\caption{Overlap of top-3 correct patches generated by context-aware learning based approaches.}
	\label{fig:venn}
\end{figure}

Figure \ref{fig:venn} demonstrates an illustration of the overlapping sets of top-3 correct patches generated by the repair techniques. Each row in this figure represents a specific approach which is color encoded for differentiability. The numbers on the right depict the total number of patches generated correctly by each approach. For example, \toolname can fix a total of 4,781 out of 11,397 bugs, \coconut can fix 3,180 out of 11,397 bugs correctly and so on. The numbers at the bottom indicate the intersecting set of the correct patches for the specific block. For example, 888 patches of the same set of bugs were correctly generated by \toolname, \coconut, and \sequencer, as shown at the bottom. As three approaches overlap, this number is reflected at the top row in this column of 888 patches. All five techniques could accurately fix 101 instances of bugs as indicated by the presence of all colors in the leftmost first column in Figure~\ref{fig:venn}. We observe that \toolname can fix a wide range of bugs, and the row for \toolname encompasses patches from other approaches. Furthermore, 891 bugs could only be fixed by \toolname as shown in the figure.

\subsection{Slicing-based context (RQ3)}
\label{slice-context-rq}
We carry out a quantitative analysis of our slicing technique to measure the type and amount of information considered as context and compare it against other approaches.

\begin{figure}[h]
	\centering
	\includegraphics[width=150pt]{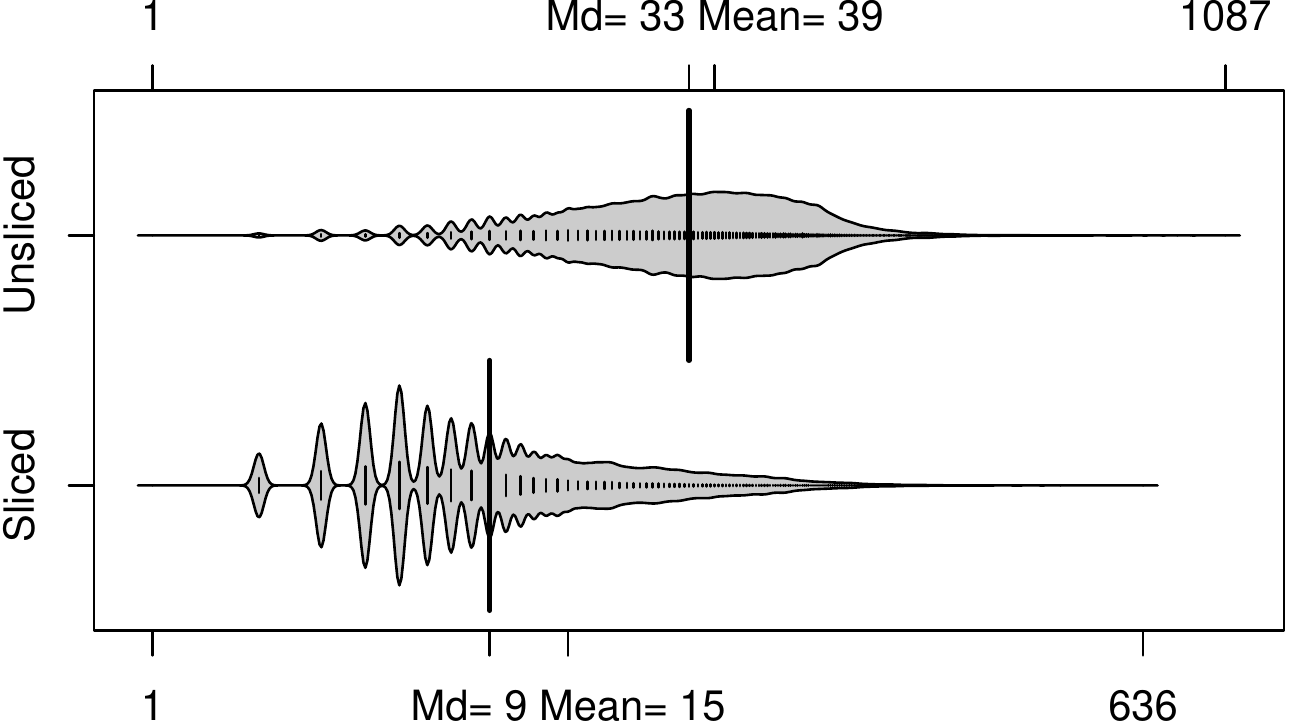}
	\caption{Distribution of number of lines in Sliced and Unsliced data.}
	\label{fig:boxplot}
\end{figure}

We collected statistics over our dataset during the program slicing analysis. We calculated the number of lines before and after slicing as well as the fraction of the datapoints undergoing control flow and data flow analysis. We found that 26.96\% of the total datapoints (113,975 pairs of buggy and fixed files) required the use of both control flow and data flow analysis to produce the slices. The remaining 73.04\% were sliced by data flow analysis only. Figure \ref{fig:boxplot} illustrates the number of lines before and after slicing across the datapoints. Here, we refer to the dataset before slicing as unsliced. 

\toolname's program slicing reduced the average number of lines to consider for context  from 39 to 15. The median number is also reduced from 33 to 9. The maximum number of lines before slicing is 1,087, and after slicing it is reduced to 636, whereas the minimum is 1 in both cases.

\begin{figure}[t]
	\centering
	\includegraphics[width=150pt]{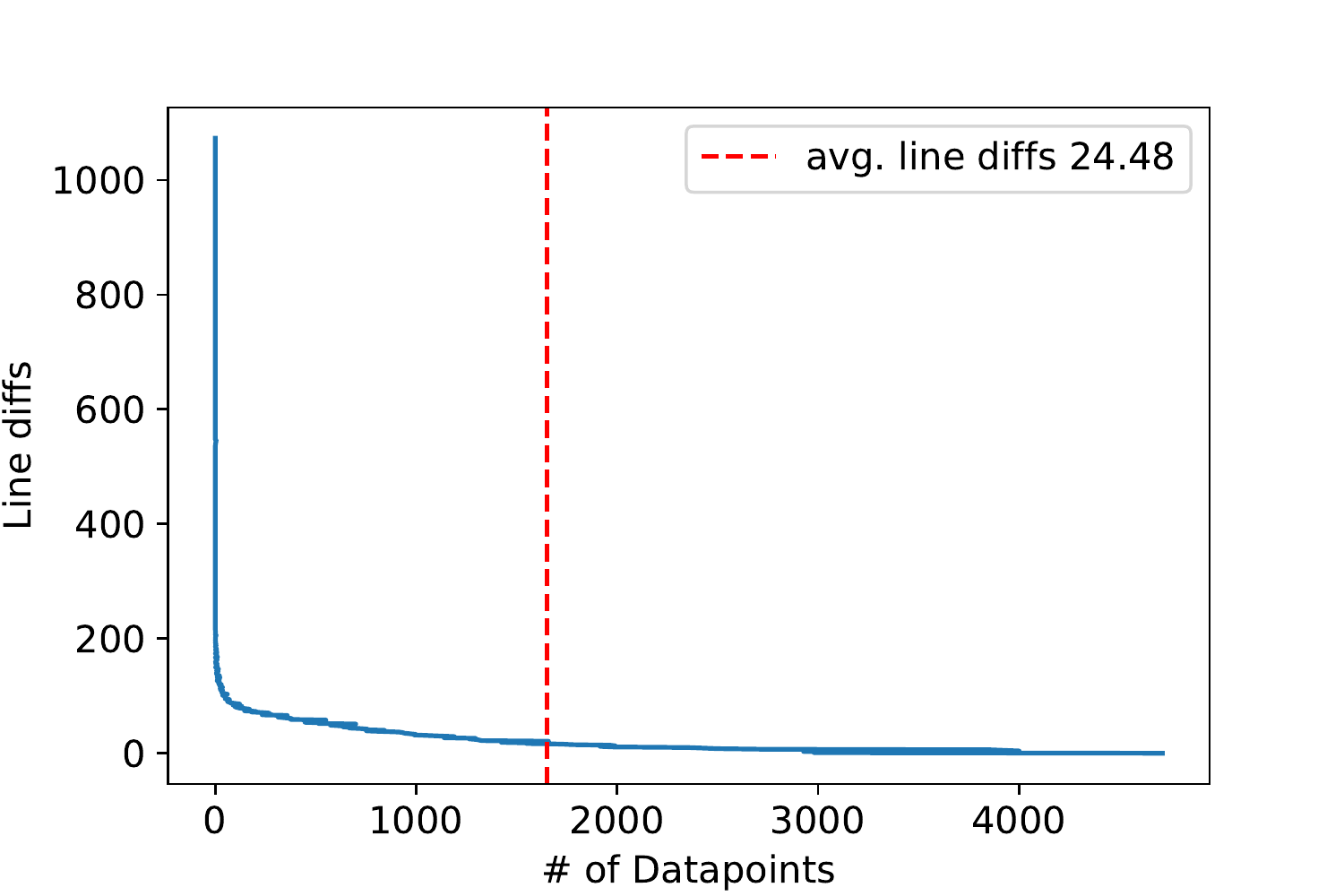}
	\caption{Reduction in number of lines after slicing across datapoints.}
	\label{fig:linediff}
\end{figure}

In Figure \ref{fig:linediff}, we can observe the spread of line differences across the datapoints. The difference from sliced to unsliced is always positive indicating reduction after slicing. The average reduction is 24 lines. The maximum reduction is 1,073 lines, for a file that was 1,087 lines.

\begin{figure*}
	\centering
	\begin{subfigure}[b]{0.15\textwidth}
		\centering
		\includegraphics[width=\textwidth,height=150pt]{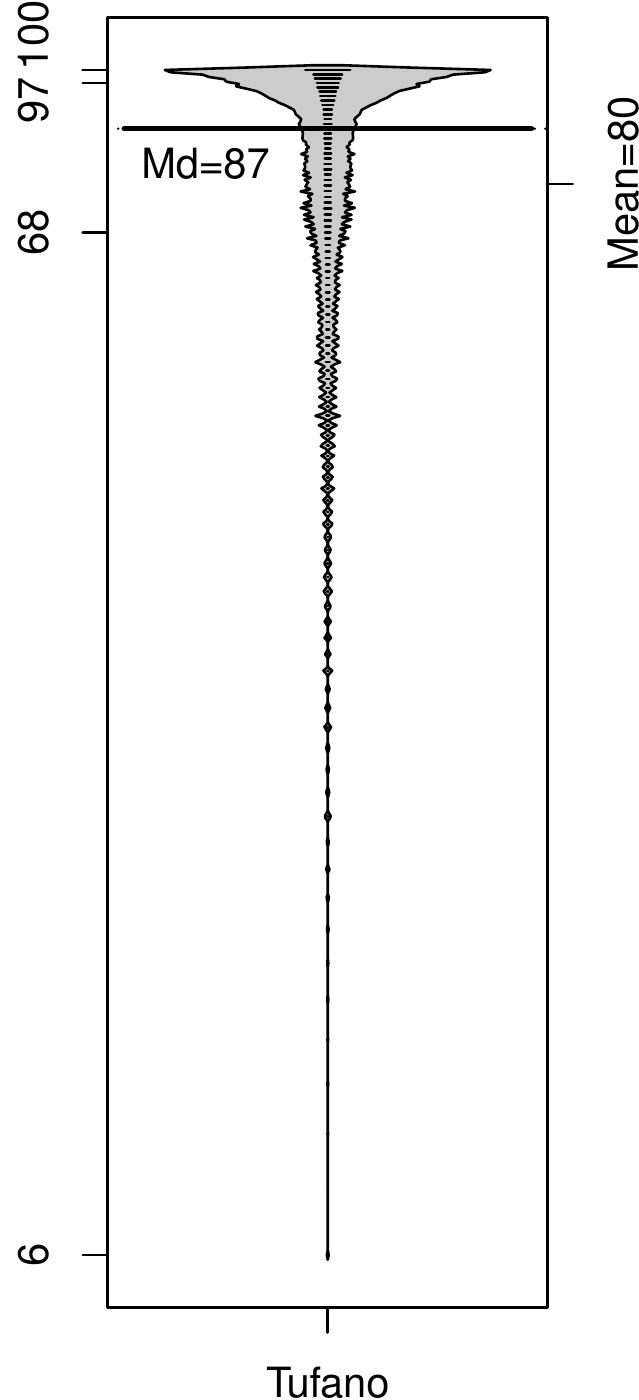}
		\label{fig:tufanotokens}
	\end{subfigure}
	\begin{subfigure}[b]{0.15\textwidth}
		\centering
		\includegraphics[width=\textwidth,height=150pt]{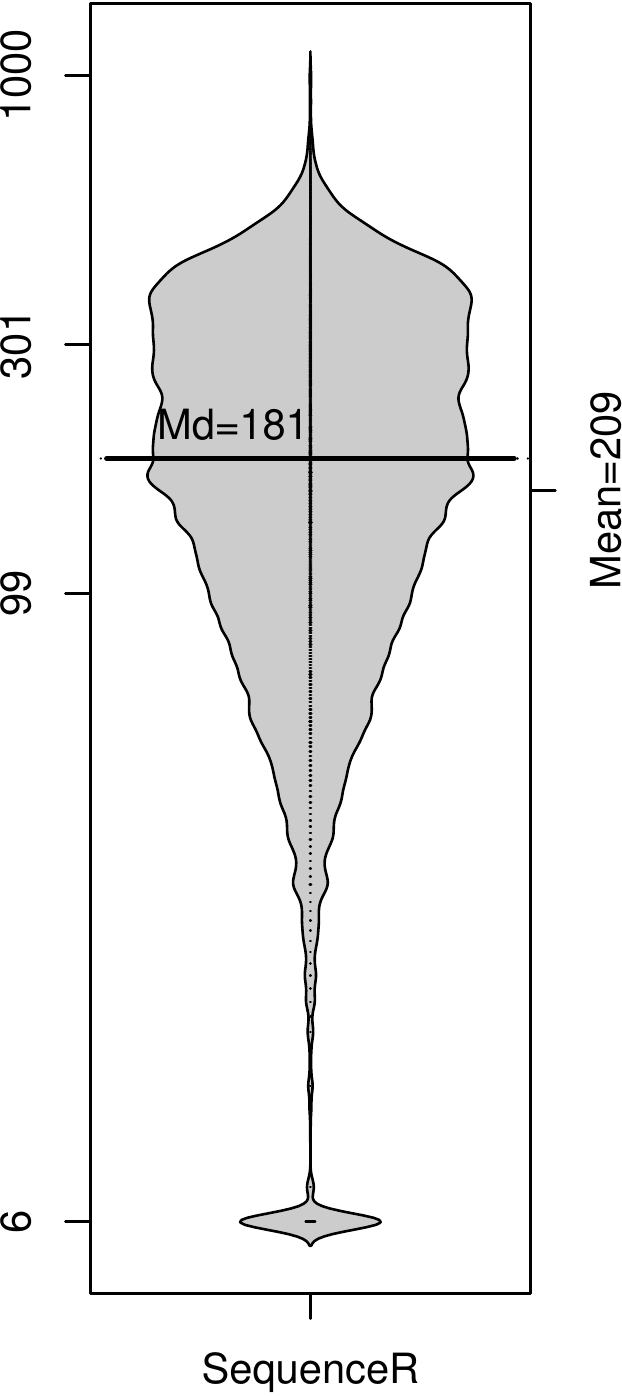}
		\label{fig:sequencertokens}
	\end{subfigure}
	\begin{subfigure}[b]{0.15\textwidth}
		\centering
		\includegraphics[width=\textwidth,height=150pt]{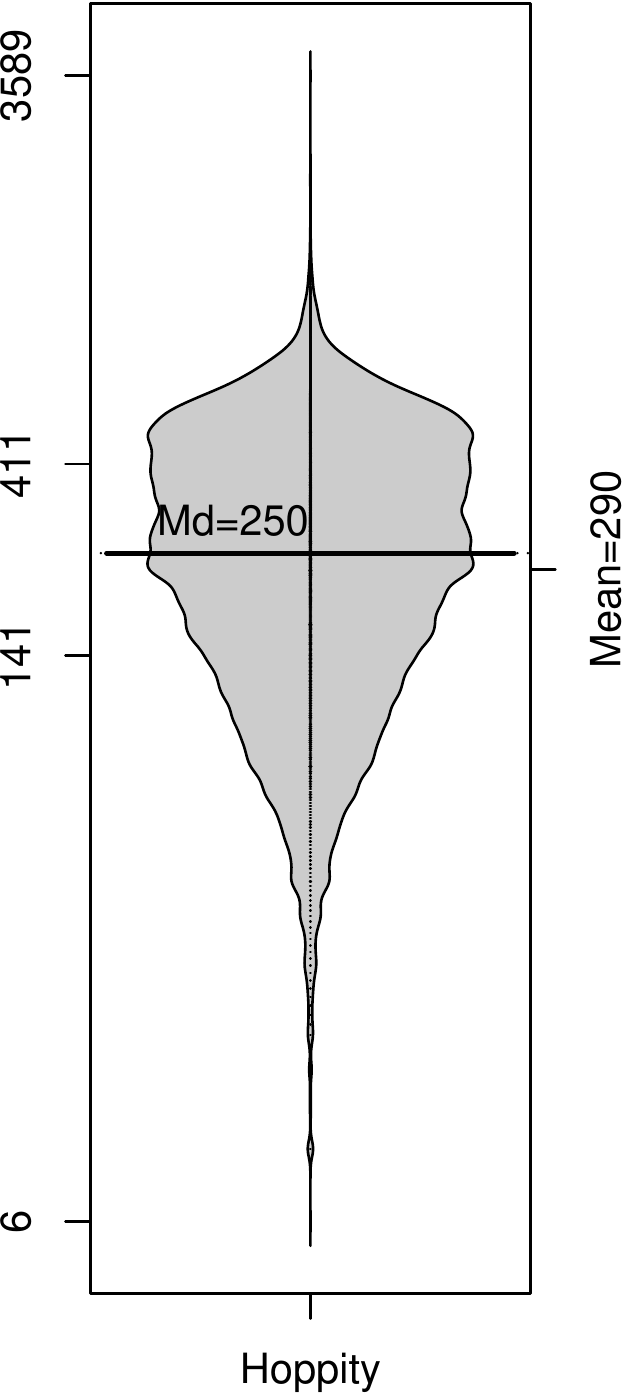}
		\label{fig:hoppitytokens}
	\end{subfigure}
	\begin{subfigure}[b]{0.15\textwidth}
		\centering
		\includegraphics[width=\textwidth,height=150pt]{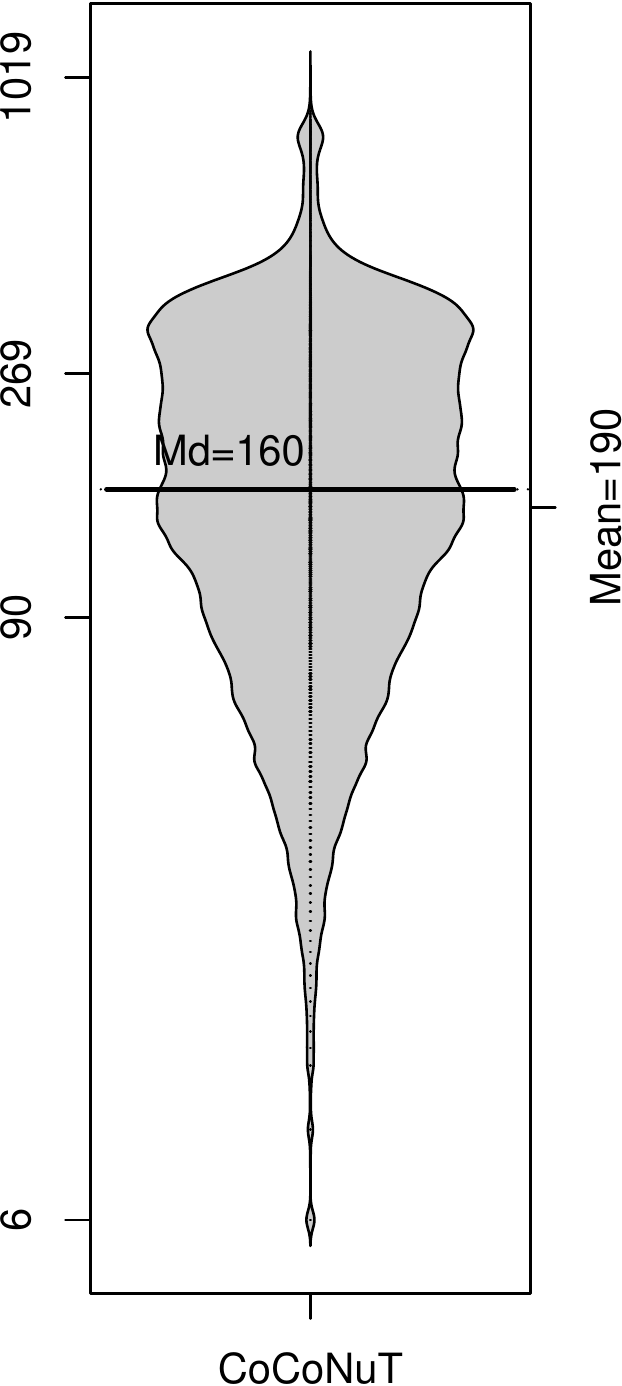}
		\label{fig:coconuttokens}
	\end{subfigure}
	\begin{subfigure}[b]{0.15\textwidth}
		\centering
		\includegraphics[width=\textwidth,height=150pt]{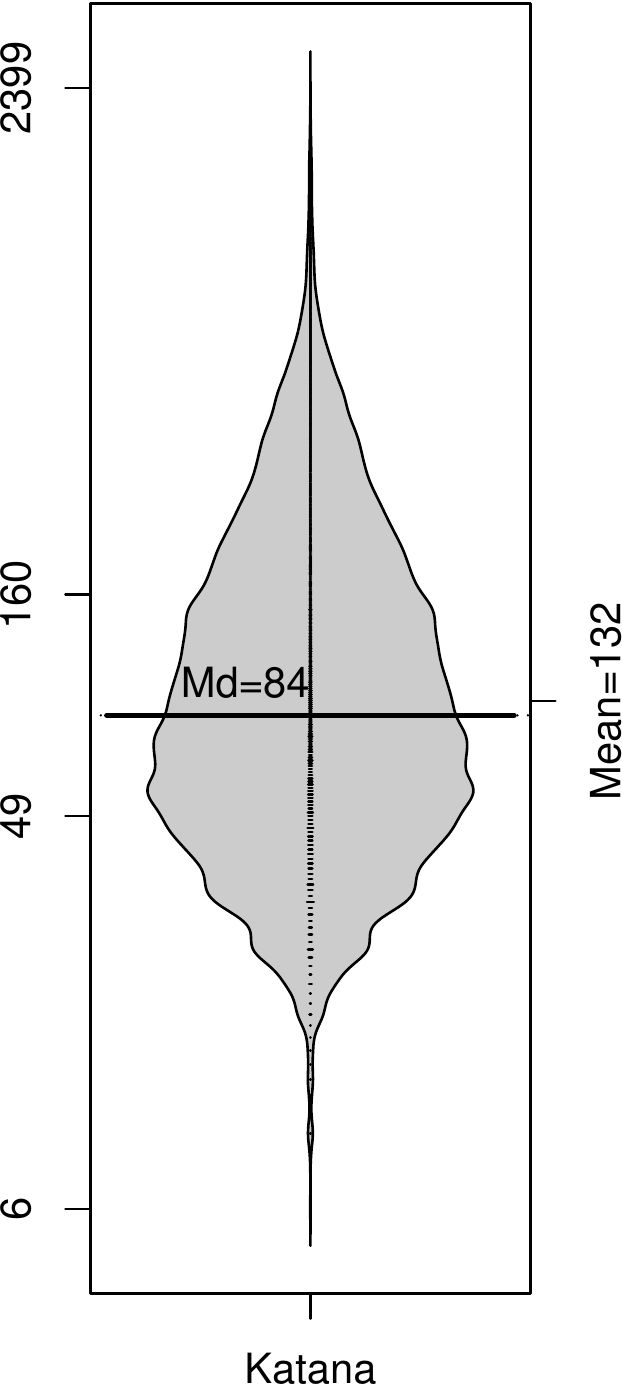}
		\label{fig:katanatokens}
	\end{subfigure}
	\caption{Descriptive statistics of tokens in context-aware learning approaches.}
	\label{fig:tokensdistro}
\end{figure*}

Figure \ref{fig:tokensdistro} depicts bean plots of the number of tokens in the buggy files for \toolname and the other techniques. We chose the buggy files from our (113,975)  JavaScript datapoints because the machine learning model takes the buggy file or buggy line with context as the input during inference. We stripped the whitespaces and comments from these files before token analysis. Tokens are the smallest possible unit that is common between all these approaches. Hence, we examine the increase or reduction of information used as context through this metric. As discussed before, except for \toolname, all of these context-based approaches use an ad-hoc bounding limit for reducing noise in the context. Just to recap, \hoppity uses a maximum limit of 500 nodes whereas \sequencer, Tufano et al. and \coconut use a sequence limit of 1,000, 100 and 1,022 tokens, respectively. Among all the approaches, \hoppity has the highest maximum token count of 3,589 tokens followed by \coconut (1,019 tokens), \sequencer (1,000 tokens) and Tufano et al. (100 tokens). The maximum number of tokens in \toolname is 2,399 tokens. Our slicing reduced the number of tokens by 1.5 times compared to \hoppity. \hoppity has the highest average number of 290 tokens, and with abstraction, Tufano et al. have the lowest average of 80 tokens. The average number of tokens in \sequencer and \coconut are 209 and 190, respectively. \toolname has an average of 132 tokens, which is less than \hoppity, \sequencer and \coconut. The median number of tokens shows a more profound effect of program slicing in the reduction of tokens. \toolname has the lowest median of 84 tokens among all the approaches. We can also observe the same trend of token reduction in the lower quartile of the dataset with \toolname. For the third quartile of the dataset, \hoppity, \sequencer and \coconut have 411, 301, and 269 tokens, respectively. On the other hand, Tufano et al. and \toolname yielded 68 and 160 tokens, which is significantly lower. Without a need for truncating the context, our approach for extracting relevant context based on control and data flow analysis, has a significant amount of information reduction on average.

\section{Discussion}

In this section, we discuss the findings from Section~\ref{sec:evaluation}.

\subsection{Role of slicing in learning repairs} In \autoref{table:rq3}, we can observe that \coconut outperforms other state-of-the-art program repair techniques. We believe that the reason for \coconut outperforming the other techniques could be attributed to using contextual information from the enclosing method, adoption of a larger model, and abstraction. \coconut utilizes two separate encoders to encode the buggy lines and surrounding context. It uses abstraction to replace all strings and numbers (except 0 and 1) with \$STRING\$ and \$NUMBER\$ to avoid out-of-vocabulary tokens. This combination of abstraction, adopting a large model, and leveraging contextual information yields better accuracy for \coconut. However, our approach \toolname outperforms \coconut by a significant margin in the top-3 and top-5 accuracy. The overall accuracy of \toolname in repairing programs can mainly be attributed to the effectiveness of dual slicing-based context. Dual slicing can derive repair ingredients from both the buggy and fixed graphs. During training, the model generates a global value dictionary containing the top 5,000 most frequent tokens. In the inference step, the model leverages this vocabulary dictionary to predict the node value for graph edit operations e.g. \code{ADD\_NODE} or \code{REP\_VAL}. The model is able to find recurring patterns for node addition or replacement in the test set, despite having only the context of only the buggy line during inference. This is because the model has encountered similar cases of node addition or node value replacement in buggy graphs during the training phase. During single slicing, the model also captures tokens from buggy and fixed graphs. However, the context is same for both buggy and fixed graphs which affects the frequency of tokens in the vocabulary. Therefore, the global value dictionary does not contain any additional tokens that may have been possible if the context from fixed line was also considered. As demonstrated in Table~\ref{table:rq1}, dual slicing is able to outperform single slicing by a significant margin due the extra contextual information. For instance, in Listing~\ref{lst:dualfix} the availability of vocabulary from the sliced context in fixed graphs during training equipped the model with repair ingredients needed for generating fixes during inference. Here, the fixed line replaces the variable \code{object} with \code{item} in which, the variable \code{item} is part of the vocabulary.

Listing~\ref{lst:slicefix} shows an example of a bug pattern that was fixed exclusively by \toolname. As shown, the fix for this bug was to add the argument \code{options} in the method call. In the sequence-based approaches, although the definition of variable \code{options} was present, the method definition of \code{generateInvite} was removed from the input sequence due to their maximum token limit. The reason is that the buggy file for this datapoint contains 435 lines of code which translates to 2,113 tokens. Although the method was defined within the enclosing class and enclosing file, the low proximity between the method call and method definition caused this method definition to be truncated. The method signature provides useful debugging information that maybe crucial to fix the bug. Through slicing, \toolname was able to retain relevant information pertaining to the buggy and fixed line. Moreover, the node value \code{options} is a commonly occuring token that was present in the global vocabulary dictionary. This indicates that dual slicing-based context can preserve repair ingredients, extract relevant information (e.g., method definition) while also reducing noise, which can benefit the learning process of the model, as shown in \autoref{table:rq3}.

\begin{lstlisting}[backgroundcolor=\color{white},frame=1,frame=single,framesep=5pt,backgroundcolor=\color{lightgray},
	rulecolor=\color{lightgray}, numbers=none,label={lst:slicefix},caption={Example of a bug fixed by \toolname}]
-   this.generateInvite();
+   this.generateInvite(options);
\end{lstlisting}

In addition to preserving the repair ingredients, program slicing can help to minimize noise for the GNN model. The quantitative analysis in Section \ref{slice-context-rq} shows the effect of program slicing in reducing the information scope, without the need for a predefined bound on the number of tokens or AST nodes. Dual slicing allows the model to focus only on the relevant context while also reducing the distance with buggy line or patch.

\begin{landscape}
	\begin{table}[!h]
		\scriptsize
		\centering
		\caption{Qualitative analysis of bugs fixed exclusively by \toolname.}
		{
\begin{tabular}{p{2.6cm}p{3.2cm}p{2cm}p{5.2cm}p{5.2cm}}
	\toprule
	\textbf{Bug Pattern} & \textbf{Description} &   \textbf{Graph Edit} & \textbf{Buggy Code} & \textbf{Fixed Code} \\
	\midrule
	Same Function More Args & \emph{Function with missing argument in the buggy line} &  \code{ADD\_NODE} & \lstinline|return drones[droneUpdate.id] = new Drone(droneUpdate);| & \lstinline|return drones[droneUpdate.id] = new Drone(droneUpdate, 10);| \\
	\cmidrule[0.3pt]{4-5}
	
	&   &  & \lstinline|currentUser: service();| & \lstinline|currentUser: service(user);|  \\
	\cmidrule[0.4pt]{1-5}
	
	Same Function Less Args & \emph{Function with additional argument in the buggy line} &  \code{DEL\_NODE}  & \lstinline|kittens.pop(name)| & \lstinline|kittens.pop()|  \\
	\cmidrule[0.3pt]{4-5}
	
	&  &  & \lstinline|app.get('/api/current_user', (req, res, next) => { | & \lstinline|app.get('/api/current_user', (req, res) => { |  \\
			\cmidrule[0.4pt]{1-5}
			
			Incorrect Object Instantiation & \emph{Constructor invoked without} \lstinline|new| \emph{keyword} &  \code{REP\_TYPE}  & \lstinline|var componentSchema = Schema({ | & \lstinline|var componentSchema = new Schema({ |  \\
					\cmidrule[0.3pt]{4-5}
					
					&  &  & \lstinline|const history = mongoose.Schema({ | & \lstinline|const history = new mongoose.Schema({ |  \\
							\cmidrule[0.4pt]{1-5}
							
							Missing Return Statement & \emph{Expression with missing}  \lstinline|return| \emph{keyword}  &  \code{REP\_TYPE} & \lstinline|await bcrypt.compare(password, savedPassword);| & \lstinline|return await bcrypt.compare(password, savedPassword);|  \\
							\cmidrule[0.3pt]{4-5}
							
							&  &  & \lstinline|changeAnimal();| & \lstinline|return changeAnimal();|  \\
							\cmidrule[0.4pt]{1-5}
							
							Incorrect Variable Declaration & \emph{Variable declared with incorrect keyword} &  \code{REP\_TYPE}  & \lstinline|var shuffled = arr.slice(0), i = arr.length, min = i - count, temp, index;| & \lstinline|let shuffled = arr.slice(0), i = arr.length, min = i - count, temp, index;|  \\
							\cmidrule[0.3pt]{4-5}
							
							&  &  & \lstinline|var garray = array//.shift()| & \lstinline|const garray = array//.shift()|  \\
							
							\cmidrule[0.4pt]{1-5}
							
							Incorrect For Loop Used & \emph{Expression with incorrect usage of} \lstinline|for...in| \emph{or}  \lstinline|for...of| \emph{used} &  \code{REP\_TYPE}  & \lstinline|for (var file in files) {| & \lstinline|for (var file of files) {| \\
									\cmidrule[0.3pt]{4-5}			 
									
							&  &  & \lstinline|for (let slot of Object.values(character.slots)) {| & \lstinline|for (let slot in Object.values(character.slots)) {| \\
									\cmidrule[0.4pt]{1-5}
									
									Change Boolean Literal & \emph{Expression with incorrect boolean literal used} &  \code{REP\_VAL} & \lstinline|app.use(bodyParser.urlencoded({ extended: false }));| & \lstinline|app.use(bodyParser.urlencoded({ extended: true }));|  \\
									\cmidrule[0.3pt]{4-5}
									
									&  &  & \lstinline|overlap: {type: Boolean, default: true},| & \lstinline|overlap: {type: Boolean, default: false},|  \\
									\cmidrule[0.4pt]{1-5}
									
									Change String/Numeric Literal & \emph{Expression with incorrect string/numeric literal used} &  \code{REP\_VAL} & \lstinline!const PORT = process.env.PORT || 3001;! & \lstinline!const PORT = process.env.PORT || 3000;!  \\
									\cmidrule[0.3pt]{4-5}
									
									&  &  & \lstinline|router.get('/comment/:id', CommentController.GetComments);| & \lstinline|router.get('/comment/list', CommentController.GetComments);| \\
									\cmidrule[0.4pt]{1-5}
									
									Change Identifier Used & \emph{Expression with incorrect identifier used} &  \code{REP\_VAL} & \lstinline|componentWillMount: () => {| & \lstinline|componentDidMount: () => {| \\
											\cmidrule[0.3pt]{4-5}			 
											
											&  &  & \lstinline|this.destroy(res, res)| & \lstinline|this.destroy(res, err)| \\
											
											\bottomrule
										\end{tabular}
									}
									\label{table:patch_analysis}
								\end{table}
							\end{landscape}

\subsection{\toolname is effective across pervasive bug patterns} Table \ref{table:patch_analysis} shows a qualitative analysis of the type and description of bug patterns observed in the test dataset that only \toolname was able to fix. Unlike prior techniques~\cite{pradeloopsla, li:suspicious-return-stmnt:20, scott:detecting-swapped-arguments:20} which are trained on specific error types, our model targets a wide variety of error types as it is trained on real world bugs. We assessed 100 random samples out of the 891 correct patches generated uniquely by  \toolname (see Figure \ref{fig:venn}, second column from right). As mentioned previously, our dataset has been curated from open-source GitHub repositories and hence, they are representative of real bugs that have been fixed by developers. For the qualitative analysis of the patches, the authors manually analyzed the resulting correct patches of \toolname, categorized them individually and came to a consensus if they agreed on the bug fixes to be representative of bug patterns. The assessment of the patches and categorization took approximately 12 person-hours in total.
For labelling the bug type, we used ManySStuBs4J~\cite{minesstubs} to identify the category of the bug fixes. As this bug type categorization is a manual process and unfeasible to perform for the entire test dataset comprising of 11,397 datapoints, we only annotated a random sample of bugs with correct predictions by \toolname. However, Table~\ref{table:data_split} demonstrates the graph edit type for the overall dataset, which has been detected automatically. We present nine bug patterns (two patches per pattern as examples) fixed by \toolname in Table \ref{table:patch_analysis}  in the test dataset. Among these nine bug patterns, five represent some common single statement bugs seen in other programming languages such as Java or Python~\cite{minesstubs,PySStuBs}. The remaining four patterns are specific to JavaScript. 
These language-specific bug patterns include  \emph{Incorrect Object Instantiation}, \emph{Missing Return Statement}, \emph{Incorrect Variable Declaration} and \emph{Incorrect For Loop Used}. Since JavaScript code is interpreted at runtime, bugs such as \emph{Incorrect Object Instantiation} and \emph{Missing Return Statement} may only surface during program execution. 
These bugs occur because JavaScript is a dynamically typed language and can silently propagate these errors. \emph{Missing Return Statement} has been categorized as a common bug pattern in JavaScript in BugsJS~\cite{bugsjs}. \emph{Incorrect Variable Instantiation} is a common best practice in modern JavaScript projects that supports the ECMAScript 6 (ES6) syntax, and it indicates a code smell. In ES6, the keywords \code{let} and \code{const} were introduced to the language \cite{es6} because of scoping issues caused by the keyword \code{var}, which can eventually lead to hard-to-find bugs in JavaScript~\cite{jsScoping}. Another bug type, \emph{Incorrect For Loop Used}, occurs mainly in ES6 compatible JavaScript source code, where the \code{for...in}  and \code{for...of} loops are incorrectly used in which the former is used for looping through enumerables in objects and the latter is used for iterating through values of arrays and objects. Due to their syntactical similarity, developers often tend to use them interchangeably for the wrong purpose which can lead to bugs that are hard to identify. As shown in \autoref{table:patch_analysis}, the bug types \emph{Same function more args} or \emph{Same function less args} use the graph edit operations \code{ADD\_NODE} and \code{DEL\_NODE}, respectively. In such cases during inference, the model only generates the patch for the buggy line by adding/deleting the identifier or literal. In case node addition, if the given node exists in the vocabulary, then the model is able to identify the node element addition. Although the model was trained on dual sliced context from both buggy and fixed files, which can sometimes introduce new context or eliminate unnecessary context pertaining to the fix, it only localizes the buggy line and generates a corresponding patch during inference. Such instances of fixed graphs with varying context, during training, does not mislead the model and as such, it is able to identify these recurring patterns to find only the buggy line and generate patch. The most recurring pattern of bugs that \toolname could fix is \emph{Change Identifier Used} which uses the graph edit operation \code{REP\_VAL} (as exhibited in Listing~\ref{lst:dualfix} and \autoref{table:patch_analysis}). Compared to other state-of-the-art tools, \toolname is most effective in fixing \emph{Change Identifier Used} bugs followed by \emph{Same function more args} and \emph{Same function less args}. This is because the extra vocabulary tokens obtained from the fixed graphs during training has equipped the model with the ability to identify and repair these types of bugs. Thus, \toolname can detect and fix a wide variety of pervasive bug patterns that include both common and language-specific bugs more effectively than the baselines.

\subsection{Effect of missing repair ingredients on inference} 
\label{subsec:ingredients}
\toolname is most effective when it can incorporate relevant repair ingredients by leveraging context using slicing. To understand why \toolname was unable to correctly generate patches for some bugs, we manually examined 200 random samples from the incorrect patches. Upon careful investigation, we observed that for 16\% of the incorrect patches,  the relevant repair ingredients were not present within the file. Since we collect our datapoints based on buggy commits, our scope is limited to the enclosing file, which often does not include  interdependent files. As such, we found instances of function calls or variables used in the buggy file that were imported from other files.  As a result, relevant syntactic and semantic information required for such functions or variables was not preserved as context for the model to generate correct patches. To mitigate this, we plan to extend our slicing technique to perform interprocedural analysis \emph{across files} as part of future work.

Furthermore, there were recurring instances (84\%) of the \emph{Change String/Numeric Literal} bug pattern among the incorrect patches generated by \toolname. Upon closer inspection, we noticed that not all of these string replacements were actual bug fixes---most of them were typo fixes or ad-hoc textual changes in config files. We believe that bug patterns pertaining to string changes can further benefit from a larger vocabulary size than the current 5,002 tokens.
\begin{lstlisting}[backgroundcolor=\color{white},frame=1,label={lst:tufanopatch},frame=single,framesep=5pt,backgroundcolor=\color{lightgray},
	rulecolor=\color{lightgray}, numbers=none,caption={Example of a bug exclusively fixed by Tufano et al.~\cite{tufano:tosem:19}.}]
-  return React . createElement ( STRING_6 , { className : STRING_7 } )
+  return React . createElement ( STRING_6 , { className : STRING_9 } )
\end{lstlisting}

However, as shown in Figure~\ref{fig:venn}, \toolname was not able to fix 420 instances of bugs that Tufano et al.~\cite{tufano:tosem:19} alone would fix. Through random sampling, we identified most of the bug patterns to be string/literal changes among these 420 bug instances. The approach used by Tufano et al.~\cite{tufano:tosem:19} uses a renaming based abstraction~\cite{controlledAPR} to limit the vocabulary size. The reason for this type of abstraction is that NMT models are not as effective in learning meaningful translations using a large vocabulary size; therefore, their approach applies abstraction to replace non-frequent identifiers and string/numeric literals with reusable tokens in the form of CATEGORY\_\#. For example, STRING\_1, NUMBER\_1, VARIABLE\_1 etc. denotes the category (identifier/literal) and the number represents the sequential occurence of that category within the code. Due to this abstraction, the search space during beam search decoding is significantly narrower in comparison to \toolname. For instance, Listing~\ref{lst:tufanopatch} depicts an example of a bug among the 420 exclusive correct patches by Tufano et al.~\cite{tufano:tosem:19}. As shown, the buggy string is replaced with the correct string which is present in the vocabulary of their model. However, the same string is not available in the vocabulary of \toolname as it is a non-frequent token. Therefore, \toolname is unable to predict the correct patch for such cases.

\begin{lstlisting}[backgroundcolor=\color{white},frame=1,label={lst:stringchangebug},frame=single,framesep=5pt,backgroundcolor=\color{lightgray},
	rulecolor=\color{lightgray}, numbers=none,caption={Example of an adhoc string literal change bug}]
-  return "foo";
+  return "bar";
\end{lstlisting}

Dual slicing by \toolname is not as effective for certain cases of bug types \emph{Change String/Numeric Literal}. These string/numeric literal changes (such as \autoref{lst:stringchangebug}) are merely typo fixes, or ad-hoc textual changes. For instance, in Section~\ref{sec:rq1} (RQ1), we notice that although dual slicing fixed 797 more bugs than single slicing, six bugs were fixed exclusively by single slicing. After assessing these six bugs, we identified that they belong to the \emph{Change String/Numeric Literal} pattern and were ad-hoc string changes. While dual slicing was able to predict the graph edit operation for these bugs (rep\_val), the value of changed string was not predicted correctly. To extract context, we use the immediate enclosing parent (i.e., file, method or class) as context, particularly when there are no variables or function calls in that line. Unless the changed string/number is present in the vocabulary dictionary, contextual information does not play an important role in program repair. Another scenario where dual slicing is not as effective is when the repair ingredients are not available within the file. This happens when an imported variable/function is used in the buggy line. For example, in Listing~\ref{lst:slicefix}, we show that \toolname was able to fix the bug as the function \code{generateInvite} was declared within the file scope, however, if it were an imported function, then the model would not have any information about the method signature (i.e., how many parameters it expects). An interprocedural analysis that spans across files could be effective in capturing this additional context.

Currently, we capture the context of all variables/functions used in the buggy line. In some cases, context of one or more variables within a buggy line may be unnecessary to fix the bug. To handle such cases, one might potentially employ code instrumentation to capture relevant information from execution traces. However, steps to perform code instrumentation and extracting runtime traces could be potentially expensive. The reason is that it requires setting up the whole project, resolving the dependencies, and making the project compilable before executing the concerned buggy line. Furthermore, for a learning-based model to be effective, a large dataset of many such examples would be required. We instead rely on contextual information that could be derived statically. As part of future work, we plan to explore code instrumentation and dynamic analysis to curate a more accurate slice.

\subsection{Limitations} 

While \toolname is a research prototype, it could be envisioned as a plugin in a developer’s IDE to facilitate program repair. For a given buggy code snippet, program slicing takes approximately 330 milliseconds. After the slicing step, the average execution time to generate a patch is 2.54 seconds, which makes it feasible to employ \toolname as a neural developer assistance tool.

To productize, \toolname could be extended by adding support for JavaScript files that contain templated code (i.e., HTML or JSX\footnote{https://reactjs.org/docs/introducing-jsx.html}) from front-end frameworks such as ReactJS\footnote{https://reactjs.org}, VueJS\footnote{https://vuejs.org} or Angular\footnote{https://angular.io}. The current implementation of our analysis tool cannot extract slices for bugs inside front-end templated code. Consequently, \toolname could produce invalid code slices because the files do not contain pure JavaScript code.

\toolname cannot currently generate slices from minified or obfuscated JavaScript code because of constraints in our slicing framework. The minification step removes whitespaces, obfuscates variable names and often produces a single line as an output.  Therefore, we excluded minified JavaScript files in our dataset as they may introduce noise into the learning process. However, in practice, this minification step is taken prior to deployment in the production environment, and it is not for human consumption. As a result, minified or obfuscated code would not limit the applicability of \toolname.

JavaScript is a dynamically typed language, which inherently makes it difficult to keep track of control and data flow dependencies during program slicing. In cases where control and data flow analysis does not yield any sliced context, we consider the contents of entire file as context scope for the buggy line. 

The current implementation of \toolname does not support multi-line patch generation. Since our dataset contains only one AST node difference, we use a single line as the slicing criterion to extract contextual statements. However, conceptually, we could extend the notion of program slicing for multi-line bug fixes.

We used both inter-procedural and intra-procedural backward slicing but limited the program analysis within the scope of a JavaScript file. One reason for such a choice is that our data collection is based on commits instead of projects; this approach cannot ensure that all the dependent files are present within the commit. Furthermore, because our underlying slicing framework, Understand~\cite{und}, only supports static analysis within the file, our JavaScript slicer is constrained by this limitation. However, according to a previous work~\cite{withinFile}, a significant portion of the fixing ingredients can be found in the file containing the bug. As discussed in Section~\ref{subsec:ingredients}, the ability of \toolname to fix bugs is constrained by the availability of vocabulary. If a bug fix requires replacing a node or adding a node, the node value needs to be present in the vocabulary dictionary. Increasing the vocabulary size could potentially improve the performance of \toolname in fixing bugs.

We used backward slicing analysis, which may have resulted in the omission of some relevant contextual statements during aliasing. Aliasing is a phenomenon where more than one variable refers to the same location in memory. Certain bugs can occur when updating an alias for a variable, which in turn mutates the original variable. As these bugs can occur beyond the buggy line, forward slicing can help capture the relevant contextual statements. Support for forward slicing could be incorporated into \toolname for this bug pattern in future work.

\section{Threats To Validity}

This section describes how we addressed potential limitations that may have biased our findings. 

\subsection{Internal Validity} An internal threat to validity is the accurate end-to-end replication of the comparison with state of the art techniques. Two of the techniques we compared with, namely Tufano et al. ~\cite{tufano:tosem:19}, and \sequencer applied custom abstraction mechanism on the enclosing method or class of bugs in Java source code. Following Tufano et al.~\cite{tufano:tosem:19}, we implemented the same abstraction technique for JavaScript. \sequencer derives contextual information from the enclosed class of the buggy line. However, JavaScript poses unique differences from a compiled language such as, Java. As mentioned in Section~\ref{sec:statprogslicing}, the buggy line in 85\% of the dataset is part of the global scope. Since JavaScript source code is not always enclosed in a class or function, we take the entire file as the context scope in such cases to ensure that the same datapoints are preserved for a fair comparison across all techniques. The length of the buggy or fixed source code for these techniques are bounded by the maximum number of tokens that each of these techniques support; the context limit for Tufano et al. is 100 tokens, \coconut is 1,022 tokens and \sequencer is 1,000 tokens. 

Both Tufano et al.~\cite{tufano:tosem:19} and \sequencer are sequence based learning techniques where source code is treated as a stream of tokens for buggy Java files; we therefore used their replication package and pre-processed our data by adding support for JavaScript. \coconut and \hoppity already supported JavaScript implementation, hence we mainly trained these models on our dataset. Another threat to validity is the varied vocabulary size for each of the state-of-the-art techniques. We used the best reported hyperparameters and vocabulary size for the respective tools to ensure a fair comparison.

\subsection{External Validity}
We build our dataset from open source GitHub repositories of JavaScript projects. We used specific keywords to filter buggy commits following prior work~\cite{minesstubs}. One external validity is that the search heuristics used cannot ensure that all the datapoints represent actual bug fixes as some commits may contain refactoring activity or ad-hoc code changes. Nevertheless, we were able to find nine types of bug patterns pervasive across the test set, and this search criteria is consistent with other learning-based program repair studies~\cite{coconut, hoppity, tufano:tosem:19}.

We implemented our approach for JavaScript, and our experiments cannot conclude how effective \toolname would be for other languages. However, the fundamental challenges \toolname addresses, i.e., how to retain useful contextual information for a repair task, why context selection using a naive approach is not sufficient, how to select useful information pertaining to the buggy/fixed line, are language independent and other languages could potentially benefit from slicing-based context. Static program slicing requires a slicing framework for each programming language, and building such a framework is an expensive process as it requires high development cost. Furthermore, state-of-the-art program repair techniques such as DLFix~\cite{dlfix}, \hoppity~\cite{hoppity}, TFix~\cite{tfix} and \sequencer~\cite{sequencer} also support a single language (i.e., Java or JavaScript) to evaluate the effectiveness of their approach. We therefore used a single programming language and built a slicing framework to evaluate our technique.

We considered a difference of one AST node for extracting buggy commits, which is similar to existing work~\cite{hoppity}. As a result, currently our approach can handle one-off errors. Existing learning-based program repair techniques~\cite{sequencer,tan:relifix:icse:2015,legouesNFWTSE2012,Scott2019GetafixLT} do not support multi-hunk changes (multiple buggy and fixed statements). However, conceptually \toolname could be extended to extract context from multiple buggy and fixed statements to better encode the repair ingredients, although this requires further evaluation in the future. In such approaches, using an ad-hoc approach would yield very large contextual information, severely limiting the effectiveness of a learning model. Therefore, we believe that fine-grained context extraction using \toolname would be even more relevant for the multi-hunk changes.

In this study, we proposed a novel technique, that leverages program slicing and the context from both buggy line and patch during training to fix bugs. We compared our approach with state-of-the-art learning based program repair techniques using the same dataset for training, validation and testing.  We also performed an ablation study to understand the effect of slicing by using the same model and representation. The underlying GNN model and code representation used by \toolname is the same as \hoppity. We re-trained \hoppity using our JavaScript dataset without slicing the buggy/fixed files. We tuned the hyper-parameters of the GNN model using our sliced dataset and ran the model with single sliced and dual sliced buggy/fixed files. Thus, we were able to determine the effect of slicing using this experimental setup.

Some techniques do not use contextual information at all and others consider context using a different approach. For instance, Rachet~\cite{hata2019learning} only considers the buggy line without any context, whereas DLFix~\cite{dlfix} selects the buggy subtree as context. Although we did not directly compare with these in our evaluation, DLFix~\cite{dlfix}  reported to outperform Rachet, and \coconut~\cite{coconut} reported to outperform DLFix. Since \toolname is directly compared with \coconut, we can deduce, through transitive closure, that \toolname outperforms DLFix and Rachet, although direct experiments are needed to verify this empirically. TFix~\cite{tfix} is categorized as a program repair tool for static errors~\cite{monperrus} and relies on reports from ESLint\footnote{https://eslint.org}, a popular static analysis tool for JavaScript. By generating the error reports based on the buggy code, TFix then merges the report with the buggy line along with its surrounding two lines of context to query the text-to-text transformer model (T5). As this has a dependency on these error reports from static analysis tools as context, we could not compare this approach with \toolname.

\header{Reproducibility} We have made our dataset, model, comparison framework, and \toolname's implementation available~\cite{katana} for reproducibility of the results. We further provide instructions for replicating our experimental setup.

\section{Related Work}

In this section, we describe related work on (a) how slicing is used in traditional program repair, (b) the usage of contextual information in learning-based program repair, and (c) the use of slicing on different learning-based source code processing tasks.

\subsection{Traditional Automated Program Repair} Automated program repair has received significant attention from the research community. Many APR techniques have been proposed that could be classified as search-based \cite{weimer:icse:2009, legouesNFWTSE2012, legoues:icse:2012,weimer:ase:2013, tan:relifix:icse:2015}, semantics-based \cite{nguyen:SemFix:icse:2013, ke:ase:2015}, or pattern-based \cite{liu:tbar:issta:2019, kim:icse:2013, saha-elixir-ase-1017, liu-saner-2018, liu-avatar-saner-2019, jiang-issta-2018}. From this large body of literature, CapGen~\cite{capgen} is the closest to our work, which employs program slicing for program repair. CapGen is a search-based program repair technique that leverages context based on forward and backward slicing of nodes in the AST for mutation operator selection and ingredient prioritization. However, CapGen requires domain-specific knowledge about bug and fix types, whereas our approach can learn patterns from sliced data and fix a variety of bug patterns.

Additionally, there are repair techniques that address specific classes of faults~\cite{luca:traditional-apr-lit-review:19}. For example,  fault specific techniques focus on repairing conditional statements~\cite{offutt:wrong-conditions:96,  xuan:wrong-conditions:17, durieux:wrong-conditions:16}, concurrency bugs~\cite{jin:concurrency-faults:11, jin:concurrency-faults:12}, string sanitization~\cite{alkhalaf:string-sanitization:14, yu:string-sanitization:11}, access control violations~\cite{son:access-control-violation:13}, or repair memory leaks~\cite{gao:memory-leaks:15}. FixMeUp~\cite{son:access-control-violation:13} targets access control violations in PHP web applications. FixMeUp applied inter-procedural program slicing on the data, and control dependence graphs to identify the statements that need to be guarded by an access control check. Instead, we applied program slicing for a learning-based model which is not bug specific.

\subsection{Learning-Based Program Repair} Most current learning-based repair techniques consider context in an ad-hoc manner. \hoppity~\cite{hoppity} is limited by a maximum of 500 nodes in the AST and treats the whole file as context for the buggy line. On the other hand, \sequencer~\cite{sequencer} uses the enclosing method and the class surrounding the buggy line to capture long-range dependencies between the buggy line and context. This technique is limited to 1,000 tokens, which may result in the truncation of code if the enclosing method is long, as an example. Tufano et al.~\cite{tufano:tosem:19} use buggy and fix files of small or medium-sized methods with a maximum of 50/100 tokens to learn bug fixes. They employ abstraction in the context of buggy and fixes files to mitigate the vocabulary limit. DLFix~\cite{dlfix} is a deep learning-based program repair approach that uses the context of the surrounding subtree of both the buggy line and fixed line to generate fixes using a tree-based RNN model. This context is summarized as a vector and used as weights to generate the patch for the buggy code. \coconut~\cite{coconut} uses the entire method of the buggy line as context with the buggy line and context fed as two separate inputs. In a following work, CURE~\cite{cure-program-repair-icse-2021} leverages GPT to provide contextual embedding for CoCoNuT. Recently, a pre-trained language model, CodeBERT~\cite{codebert-java-sstubs-21}, has been applied to fix Java simple bugs. There have also been attempts to generate patches in a refined way~\cite{recoder-fse-2021}, generate multi-statement fixes~\cite{dear-icse-2022} and to adjust the loss function during training iterations~\cite{rewardrepair-icse-2022}. TFix~\cite{tfix} is a learning-based system that uses a T5 model~\cite{t5-model} fine-tuned on a text-to-text patch prediction task that leverages the surrounding two lines of the buggy line as context for linting errors in JavaScript. All these techniques leverage context in a limited way and rely on predefined heuristics. In contrast, \toolname relies on program analysis to retrieve relevant code as context, is not bounded by predefined limits or heuristics, and leverages control and data flow information from both the buggy and fixed versions of the code. 

\subsection{Slicing in Learning-Based Source Code Processing} Program slicing was explored in a learning-based vulnerability detection task~\cite{li:sysevr-vulnerability-detection:21, VulDeePeckerBinary, VulDeePeckerMulti}. VulDeePecker~\cite{VulDeePeckerBinary} trains a neural network with positive and negative examples to determine whether a code snippet suffers from vulnerability. A more recent development is µVulDeePecker~\cite{VulDeePeckerMulti},  which extends VulDeePecker to predict multiclass vulnerabilities. These techniques extract slices of the vulnerable library/API function calls and are trained on an RNN to predict vulnerabilities in code. In a recent work, Xiao et al.~\cite{slicing:apirecommendation:2021} applied program slicing to a learning-based cryptographic API Suggestion. This approach applied interprocedural backward slicing from the invocation statement of a cryptography API and trained a modified LSTM-based model for a neural-network-based API recommendation task. In \toolname, we instead (a) apply program slicing on a generative task, i.e., fix generation, (b) introduce the notion of dual slicing to learn from both the buggy and fixed-versions of the code, and (c) employ a GNN model to better capture rich structural and semantic information that is inherent to the source code instead of using RNNs that treat code as a linear stream of tokens. To the best of our knowledge, we are the first to employ slicing for learning-based program repair.

\section{Conclusion}
As a developer, context is pivotal while understanding and writing a fix for a buggy piece of code. Surrounding lines, methods, and class level information provide background contextual information that is important to devise a patch for a buggy piece of code. Current learning-based repair techniques adopt predefined heuristics, such as a limited number of tokens or AST nodes to extract context, which is insufficient for representing contextual information of a bug fix. We argue that context extraction needs to be directed toward relevant code for a learning-based repair technique to be effective. We present \toolname, the first technique to apply program slicing on a learning-based program repair task that includes relevant information as context pertaining to the buggy/fixed code. We show that program slicing through control and data flow analysis effectively preserves sufficient program repair ingredients to extract recurring patterns.  We propose the notion of \emph{dual slicing}, a novel approach that leverages context from both the buggy and corrected code. Our approach is able to fix 4,781 out of 11,397 bugs curated from open-source JavaScript projects. The results show that \toolname can resolve between 1.5 up to 3.7 times more bugs when compared to the state-of-the-art learning-based repair techniques. Furthermore, \toolname is effective across a wide range of bug patterns.
In the future, we plan to expand our interprocedural analysis, which is now limited to the enclosing file. We will also expand \toolname to support other programming languages such as Java.

\bibliographystyle{ACM-Reference-Format}
\interlinepenalty=10000
\bibliography{learning-on-source-code-processing-tasks}


\begin{thebibliography}{70}


\ifx \showCODEN    \undefined \def \showCODEN     #1{\unskip}     \fi
\ifx \showDOI      \undefined \def \showDOI       #1{#1}\fi
\ifx \showISBNx    \undefined \def \showISBNx     #1{\unskip}     \fi
\ifx \showISBNxiii \undefined \def \showISBNxiii  #1{\unskip}     \fi
\ifx \showISSN     \undefined \def \showISSN      #1{\unskip}     \fi
\ifx \showLCCN     \undefined \def \showLCCN      #1{\unskip}     \fi
\ifx \shownote     \undefined \def \shownote      #1{#1}          \fi
\ifx \showarticletitle \undefined \def \showarticletitle #1{#1}   \fi
\ifx \showURL      \undefined \def \showURL       {\relax}        \fi
\providecommand\bibfield[2]{#2}
\providecommand\bibinfo[2]{#2}
\providecommand\natexlab[1]{#1}
\providecommand\showeprint[2][]{arXiv:#2}

\bibitem[\protect\citeauthoryear{??}{es6}{2015}]%
        {es6}
 \bibinfo{year}{2015}\natexlab{}.
\newblock \bibinfo{title}{ECMAScript 2015 Language Specification - ECMA-262 6th
  Edition}.
\newblock \bibinfo{howpublished}{\url{https://262.ecma-international.org/6.0}}.
\newblock
\newblock
\shownote{Accessed: 2022-01-07.}


\bibitem[\protect\citeauthoryear{??}{sta}{2021}]%
        {stackOSurvey}
 \bibinfo{year}{2021}\natexlab{}.
\newblock \bibinfo{title}{StackOverflow Developer Survey 2021}.
\newblock
  \bibinfo{howpublished}{\url{https://insights.stackoverflow.com/survey/2021/\#most-popular-technologies-language-prof}}.
\newblock
\newblock
\shownote{Accessed: 2022-01-26.}


\bibitem[\protect\citeauthoryear{??}{und}{2021}]%
        {und}
 \bibinfo{year}{2021}\natexlab{}.
\newblock \bibinfo{title}{Understand by Scitools}.
\newblock \bibinfo{howpublished}{\url{https://www.scitools.com/}}.
\newblock
\newblock
\shownote{Accessed: 2021-12-30.}


\bibitem[\protect\citeauthoryear{??}{kat}{2022}]%
        {katana}
 \bibinfo{year}{2022}\natexlab{}.
\newblock \bibinfo{title}{Katana}.
\newblock \bibinfo{howpublished}{\url{https://github.com/saltlab/Katana}}.
\newblock
\newblock
\shownote{Accessed: 2022-11-25.}


\bibitem[\protect\citeauthoryear{Alkhalaf, Aydin, and Bultan}{Alkhalaf
  et~al\mbox{.}}{2014}]%
        {alkhalaf:string-sanitization:14}
\bibfield{author}{\bibinfo{person}{Muath Alkhalaf}, \bibinfo{person}{Abdulbaki
  Aydin}, {and} \bibinfo{person}{Tevfik Bultan}.}
  \bibinfo{year}{2014}\natexlab{}.
\newblock \showarticletitle{Semantic Differential Repair for Input Validation
  and Sanitization}. In \bibinfo{booktitle}{\emph{Proceedings of the 2014
  International Symposium on Software Testing and Analysis}}
  \emph{(\bibinfo{series}{ISSTA 2014})}. \bibinfo{publisher}{Association for
  Computing Machinery}, \bibinfo{pages}{225–236}.
\newblock


\bibitem[\protect\citeauthoryear{Allamanis}{Allamanis}{2019}]%
        {allamanis2019adverse}
\bibfield{author}{\bibinfo{person}{Miltiadis Allamanis}.}
  \bibinfo{year}{2019}\natexlab{}.
\newblock \showarticletitle{The Adverse Effects of Code Duplication in Machine
  Learning Models of Code}. In \bibinfo{booktitle}{\emph{Proceedings of the
  2019 ACM SIGPLAN International Symposium on New Ideas, New Paradigms, and
  Reflections on Programming and Software}} \emph{(\bibinfo{series}{Onward!
  2019})}. \bibinfo{publisher}{Association for Computing Machinery},
  \bibinfo{pages}{143–153}.
\newblock


\bibitem[\protect\citeauthoryear{Allamanis, Brockschmidt, and
  Khademi}{Allamanis et~al\mbox{.}}{2018}]%
        {allamanis-represent-programs-with-graphs-2018}
\bibfield{author}{\bibinfo{person}{Miltiadis Allamanis}, \bibinfo{person}{Marc
  Brockschmidt}, {and} \bibinfo{person}{Mahmoud Khademi}.}
  \bibinfo{year}{2018}\natexlab{}.
\newblock \showarticletitle{Learning to Represent Programs with Graphs}. In
  \bibinfo{booktitle}{\emph{International Conference on Learning
  Representations}} \emph{(\bibinfo{series}{ICLR})}. \bibinfo{pages}{520--524}.
\newblock


\bibitem[\protect\citeauthoryear{Bader, Scott, Pradel, and Chandra}{Bader
  et~al\mbox{.}}{2019}]%
        {Scott2019GetafixLT}
\bibfield{author}{\bibinfo{person}{Johannes Bader}, \bibinfo{person}{Andrew
  Scott}, \bibinfo{person}{Michael Pradel}, {and} \bibinfo{person}{Satish
  Chandra}.} \bibinfo{year}{2019}\natexlab{}.
\newblock \showarticletitle{Getafix: Learning to Fix Bugs Automatically}.
\newblock \bibinfo{journal}{\emph{Proceedings of the ACM on Programming
  Languages}} \bibinfo{number}{OOPSLA} (\bibinfo{year}{2019}),
  \bibinfo{numpages}{27}~pages.
\newblock


\bibitem[\protect\citeauthoryear{Barr, Brun, Devanbu, Harman, and Sarro}{Barr
  et~al\mbox{.}}{2014}]%
        {withinFile}
\bibfield{author}{\bibinfo{person}{Earl~T. Barr}, \bibinfo{person}{Yuriy Brun},
  \bibinfo{person}{Premkumar Devanbu}, \bibinfo{person}{Mark Harman}, {and}
  \bibinfo{person}{Federica Sarro}.} \bibinfo{year}{2014}\natexlab{}.
\newblock \showarticletitle{The Plastic Surgery Hypothesis}
  \emph{(\bibinfo{series}{FSE '14})}. \bibinfo{publisher}{Association for
  Computing Machinery}, \bibinfo{pages}{306–317}.
\newblock


\bibitem[\protect\citeauthoryear{Berabi, He, Raychev, and Vechev}{Berabi
  et~al\mbox{.}}{2021}]%
        {tfix}
\bibfield{author}{\bibinfo{person}{Berkay Berabi}, \bibinfo{person}{Jingxuan
  He}, \bibinfo{person}{Veselin Raychev}, {and} \bibinfo{person}{Martin
  Vechev}.} \bibinfo{year}{2021}\natexlab{}.
\newblock \showarticletitle{Tfix: Learning to fix coding errors with a
  text-to-text transformer}. In \bibinfo{booktitle}{\emph{International
  Conference on Machine Learning}}. PMLR, \bibinfo{pages}{780--791}.
\newblock


\bibitem[\protect\citeauthoryear{{Chen}, {Kommrusch}, {Tufano}, {Pouchet},
  {Poshyvanyk}, and {Monperrus}}{{Chen} et~al\mbox{.}}{2019}]%
        {sequencer}
\bibfield{author}{\bibinfo{person}{Z. {Chen}}, \bibinfo{person}{S.~J.
  {Kommrusch}}, \bibinfo{person}{M. {Tufano}}, \bibinfo{person}{L. {Pouchet}},
  \bibinfo{person}{D. {Poshyvanyk}}, {and} \bibinfo{person}{M. {Monperrus}}.}
  \bibinfo{year}{2019}\natexlab{}.
\newblock \showarticletitle{SEQUENCER: Sequence-to-Sequence Learning for
  End-to-End Program Repair}.
\newblock \bibinfo{journal}{\emph{IEEE Transactions on Software Engineering}}
  (\bibinfo{year}{2019}), \bibinfo{pages}{1--1}.
\newblock


\bibitem[\protect\citeauthoryear{Dinella, Dai, Li, Naik, Song, and
  Wang}{Dinella et~al\mbox{.}}{2020}]%
        {hoppity}
\bibfield{author}{\bibinfo{person}{Elizabeth Dinella}, \bibinfo{person}{Hanjun
  Dai}, \bibinfo{person}{Ziyang Li}, \bibinfo{person}{Mayur Naik},
  \bibinfo{person}{Le Song}, {and} \bibinfo{person}{Ke Wang}.}
  \bibinfo{year}{2020}\natexlab{}.
\newblock \showarticletitle{Hoppity: Learning Graph Transformations to Detect
  and Fix Bugs in Programs}. In \bibinfo{booktitle}{\emph{International
  Conference on Learning Representations}} \emph{(\bibinfo{series}{ICLR})}.
\newblock


\bibitem[\protect\citeauthoryear{Durieux and Monperrus}{Durieux and
  Monperrus}{2016}]%
        {durieux:wrong-conditions:16}
\bibfield{author}{\bibinfo{person}{Thomas Durieux} {and}
  \bibinfo{person}{Martin Monperrus}.} \bibinfo{year}{2016}\natexlab{}.
\newblock \showarticletitle{DynaMoth: Dynamic Code Synthesis for Automatic
  Program Repair}. In \bibinfo{booktitle}{\emph{Proceedings of the 11th
  International Workshop on Automation of Software Test}} (Austin, Texas)
  \emph{(\bibinfo{series}{AST '16})}. \bibinfo{publisher}{Association for
  Computing Machinery}, \bibinfo{pages}{85–91}.
\newblock


\bibitem[\protect\citeauthoryear{Gao, Xiong, Mi, Zhang, Yang, Zhou, Xie, and
  Mei}{Gao et~al\mbox{.}}{2015}]%
        {gao:memory-leaks:15}
\bibfield{author}{\bibinfo{person}{Qing Gao}, \bibinfo{person}{Yingfei Xiong},
  \bibinfo{person}{Yaqing Mi}, \bibinfo{person}{Lu Zhang},
  \bibinfo{person}{Weikun Yang}, \bibinfo{person}{Zhaoping Zhou},
  \bibinfo{person}{Bing Xie}, {and} \bibinfo{person}{Hong Mei}.}
  \bibinfo{year}{2015}\natexlab{}.
\newblock \showarticletitle{Safe Memory-Leak Fixing for C Programs}. In
  \bibinfo{booktitle}{\emph{Proceedings of the 37th International Conference on
  Software Engineering - Volume 1}} \emph{(\bibinfo{series}{ICSE '15})}.
  \bibinfo{publisher}{IEEE Press}, \bibinfo{pages}{459–470}.
\newblock


\bibitem[\protect\citeauthoryear{Gazzola, Micucci, and Mariani}{Gazzola
  et~al\mbox{.}}{2019}]%
        {luca:traditional-apr-lit-review:19}
\bibfield{author}{\bibinfo{person}{Luca Gazzola}, \bibinfo{person}{Daniela
  Micucci}, {and} \bibinfo{person}{Leonardo Mariani}.}
  \bibinfo{year}{2019}\natexlab{}.
\newblock \showarticletitle{Automatic Software Repair: A Survey}.
\newblock \bibinfo{journal}{\emph{IEEE Transactions on Software Engineering}}
  \bibinfo{volume}{45}, \bibinfo{number}{1} (\bibinfo{year}{2019}),
  \bibinfo{pages}{34--67}.
\newblock


\bibitem[\protect\citeauthoryear{Gupta, Pal, Kanade, and Shevade}{Gupta
  et~al\mbox{.}}{2017}]%
        {DeepFixFC}
\bibfield{author}{\bibinfo{person}{Rahul Gupta}, \bibinfo{person}{Soham Pal},
  \bibinfo{person}{Aditya Kanade}, {and} \bibinfo{person}{Shirish Shevade}.}
  \bibinfo{year}{2017}\natexlab{}.
\newblock \showarticletitle{DeepFix: Fixing Common C Language Errors by Deep
  Learning}. In \bibinfo{booktitle}{\emph{Proceedings of the Thirty-First AAAI
  Conference on Artificial Intelligence}}. \bibinfo{publisher}{AAAI Press},
  \bibinfo{pages}{1345–1351}.
\newblock


\bibitem[\protect\citeauthoryear{Gupta and Gupta}{Gupta and Gupta}{2019}]%
        {noiseML}
\bibfield{author}{\bibinfo{person}{Shivani Gupta} {and} \bibinfo{person}{Atul
  Gupta}.} \bibinfo{year}{2019}\natexlab{}.
\newblock \showarticletitle{Dealing with Noise Problem in Machine Learning
  Data-sets: A Systematic Review}.
\newblock \bibinfo{journal}{\emph{Procedia Computer Science}}
  (\bibinfo{year}{2019}), \bibinfo{pages}{466--474}.
\newblock


\bibitem[\protect\citeauthoryear{Gyimesi, Vancsics, Stocco, Mazinanian,
  Beszédes, Ferenc, and Mesbah}{Gyimesi et~al\mbox{.}}{2019}]%
        {bugsjs}
\bibfield{author}{\bibinfo{person}{Péter Gyimesi}, \bibinfo{person}{Béla
  Vancsics}, \bibinfo{person}{Andrea Stocco}, \bibinfo{person}{Davood
  Mazinanian}, \bibinfo{person}{Árpád Beszédes}, \bibinfo{person}{Rudolf
  Ferenc}, {and} \bibinfo{person}{Ali Mesbah}.}
  \bibinfo{year}{2019}\natexlab{}.
\newblock \showarticletitle{BugsJS: a Benchmark of JavaScript Bugs}. In
  \bibinfo{booktitle}{\emph{2019 12th IEEE Conference on Software Testing,
  Validation and Verification (ICST)}}. \bibinfo{pages}{90--101}.
\newblock


\bibitem[\protect\citeauthoryear{Haque, LeClair, Wu, and McMillan}{Haque
  et~al\mbox{.}}{2020}]%
        {subroutineSummarization}
\bibfield{author}{\bibinfo{person}{Sakib Haque}, \bibinfo{person}{Alexander
  LeClair}, \bibinfo{person}{Lingfei Wu}, {and} \bibinfo{person}{Collin
  McMillan}.} \bibinfo{year}{2020}\natexlab{}.
\newblock \bibinfo{booktitle}{\emph{Improved Automatic Summarization of
  Subroutines via Attention to File Context}}.
\newblock \bibinfo{publisher}{Association for Computing Machinery},
  \bibinfo{pages}{300–310}.
\newblock


\bibitem[\protect\citeauthoryear{Hata, Shihab, and Neubig}{Hata
  et~al\mbox{.}}{2019}]%
        {hata2019learning}
\bibfield{author}{\bibinfo{person}{Hideaki Hata}, \bibinfo{person}{Emad
  Shihab}, {and} \bibinfo{person}{Graham Neubig}.}
  \bibinfo{year}{2019}\natexlab{}.
\newblock \showarticletitle{Learning to Generate Corrective Patches using
  Neural Machine Translation}.
\newblock \bibinfo{journal}{\emph{arXiv preprint arXiv:1812.07170}}
  (\bibinfo{year}{2019}).
\newblock


\bibitem[\protect\citeauthoryear{Jiang, Xiong, Zhang, Gao, and Chen}{Jiang
  et~al\mbox{.}}{2018}]%
        {jiang-issta-2018}
\bibfield{author}{\bibinfo{person}{Jiajun Jiang}, \bibinfo{person}{Yingfei
  Xiong}, \bibinfo{person}{Hongyu Zhang}, \bibinfo{person}{Qing Gao}, {and}
  \bibinfo{person}{Xiangqun Chen}.} \bibinfo{year}{2018}\natexlab{}.
\newblock \showarticletitle{Shaping Program Repair Space with Existing Patches
  and Similar Code}. In \bibinfo{booktitle}{\emph{Proceedings of the 27th ACM
  SIGSOFT International Symposium on Software Testing and Analysis}}
  \emph{(\bibinfo{series}{ISSTA})}. \bibinfo{publisher}{Association for
  Computing Machinery}, \bibinfo{pages}{298–309}.
\newblock


\bibitem[\protect\citeauthoryear{Jiang, Lutellier, and Tan}{Jiang
  et~al\mbox{.}}{2021}]%
        {cure-program-repair-icse-2021}
\bibfield{author}{\bibinfo{person}{N. Jiang}, \bibinfo{person}{T. Lutellier},
  {and} \bibinfo{person}{L. Tan}.} \bibinfo{year}{2021}\natexlab{}.
\newblock \showarticletitle{CURE: Code-Aware Neural Machine Translation for
  Automatic Program Repair}. In \bibinfo{booktitle}{\emph{2021 IEEE/ACM 43rd
  International Conference on Software Engineering (ICSE)}}.
  \bibinfo{publisher}{IEEE Computer Society}, \bibinfo{pages}{1161--1173}.
\newblock


\bibitem[\protect\citeauthoryear{Jin, Song, Zhang, Lu, and Liblit}{Jin
  et~al\mbox{.}}{2011}]%
        {jin:concurrency-faults:11}
\bibfield{author}{\bibinfo{person}{Guoliang Jin}, \bibinfo{person}{Linhai
  Song}, \bibinfo{person}{Wei Zhang}, \bibinfo{person}{Shan Lu}, {and}
  \bibinfo{person}{Ben Liblit}.} \bibinfo{year}{2011}\natexlab{}.
\newblock \showarticletitle{Automated Atomicity-Violation Fixing}. In
  \bibinfo{booktitle}{\emph{Proceedings of the 32nd ACM SIGPLAN Conference on
  Programming Language Design and Implementation}} \emph{(\bibinfo{series}{PLDI
  '11})}. \bibinfo{publisher}{Association for Computing Machinery},
  \bibinfo{pages}{389–400}.
\newblock


\bibitem[\protect\citeauthoryear{Jin, Zhang, Deng, Liblit, and Lu}{Jin
  et~al\mbox{.}}{2012}]%
        {jin:concurrency-faults:12}
\bibfield{author}{\bibinfo{person}{Guoliang Jin}, \bibinfo{person}{Wei Zhang},
  \bibinfo{person}{Dongdong Deng}, \bibinfo{person}{Ben Liblit}, {and}
  \bibinfo{person}{Shan Lu}.} \bibinfo{year}{2012}\natexlab{}.
\newblock \showarticletitle{Automated Concurrency-Bug Fixing}. In
  \bibinfo{booktitle}{\emph{Proceedings of the 10th USENIX Conference on
  Operating Systems Design and Implementation}}
  \emph{(\bibinfo{series}{OSDI'12})}. \bibinfo{publisher}{USENIX Association},
  \bibinfo{pages}{221–236}.
\newblock


\bibitem[\protect\citeauthoryear{Kamienski, Palechor, Bezemer, and
  Hindle}{Kamienski et~al\mbox{.}}{2021}]%
        {PySStuBs}
\bibfield{author}{\bibinfo{person}{Arthur~V. Kamienski}, \bibinfo{person}{Luisa
  Palechor}, \bibinfo{person}{Cor-Paul Bezemer}, {and} \bibinfo{person}{Abram
  Hindle}.} \bibinfo{year}{2021}\natexlab{}.
\newblock \showarticletitle{PySStuBs: Characterizing Single-Statement Bugs in
  Popular Open-Source Python Projects}. In \bibinfo{booktitle}{\emph{2021
  IEEE/ACM 18th International Conference on Mining Software Repositories
  (MSR)}}. \bibinfo{pages}{520--524}.
\newblock


\bibitem[\protect\citeauthoryear{Karampatsis and Sutton}{Karampatsis and
  Sutton}{2020}]%
        {minesstubs}
\bibfield{author}{\bibinfo{person}{Rafael-Michael Karampatsis} {and}
  \bibinfo{person}{Charles Sutton}.} \bibinfo{year}{2020}\natexlab{}.
\newblock \bibinfo{booktitle}{\emph{How Often Do Single-Statement Bugs Occur?
  The ManySStuBs4J Dataset}}.
\newblock \bibinfo{publisher}{Association for Computing Machinery},
  \bibinfo{pages}{573–577}.
\newblock


\bibitem[\protect\citeauthoryear{Ke, Stolee, Goues, and Brun}{Ke
  et~al\mbox{.}}{2015}]%
        {ke:ase:2015}
\bibfield{author}{\bibinfo{person}{Yalin Ke}, \bibinfo{person}{Kathryn~T.
  Stolee}, \bibinfo{person}{Claire~Le Goues}, {and} \bibinfo{person}{Yuriy
  Brun}.} \bibinfo{year}{2015}\natexlab{}.
\newblock \showarticletitle{Repairing Programs with Semantic Code Search}. In
  \bibinfo{booktitle}{\emph{Proceedings of the 30th IEEE/ACM International
  Conference on Automated Software Engineering}}
  \emph{(\bibinfo{series}{ASE})}. \bibinfo{publisher}{IEEE Press},
  \bibinfo{pages}{295–306}.
\newblock


\bibitem[\protect\citeauthoryear{Kim, Nam, Song, and Kim}{Kim
  et~al\mbox{.}}{2013}]%
        {kim:icse:2013}
\bibfield{author}{\bibinfo{person}{Dongsun Kim}, \bibinfo{person}{Jaechang
  Nam}, \bibinfo{person}{Jaewoo Song}, {and} \bibinfo{person}{Sunghun Kim}.}
  \bibinfo{year}{2013}\natexlab{}.
\newblock \showarticletitle{Automatic Patch Generation Learned from
  Human-Written Patches}. In \bibinfo{booktitle}{\emph{Proceedings of the 2013
  International Conference on Software Engineering}}
  \emph{(\bibinfo{series}{ICSE})}. \bibinfo{publisher}{IEEE Press},
  \bibinfo{pages}{802–811}.
\newblock


\bibitem[\protect\citeauthoryear{Kim and Kim}{Kim and Kim}{2019}]%
        {confix}
\bibfield{author}{\bibinfo{person}{Jindae Kim} {and} \bibinfo{person}{Sunghun
  Kim}.} \bibinfo{year}{2019}\natexlab{}.
\newblock \showarticletitle{Automatic patch generation with context-based
  change application}.
\newblock \bibinfo{journal}{\emph{Empirical Software Engineering}}
  (\bibinfo{year}{2019}), \bibinfo{pages}{4071--4106}.
\newblock


\bibitem[\protect\citeauthoryear{Ko, Myers, Coblenz, and Aung}{Ko
  et~al\mbox{.}}{2006}]%
        {developersBugs}
\bibfield{author}{\bibinfo{person}{Andrew~J. Ko}, \bibinfo{person}{Brad~A.
  Myers}, \bibinfo{person}{Michael~J. Coblenz}, {and}
  \bibinfo{person}{Htet~Htet Aung}.} \bibinfo{year}{2006}\natexlab{}.
\newblock \showarticletitle{An Exploratory Study of How Developers Seek,
  Relate, and Collect Relevant Information during Software Maintenance Tasks}.
\newblock \bibinfo{journal}{\emph{IEEE Trans. Softw. Eng.}}
  \bibinfo{volume}{32}, \bibinfo{number}{12} (\bibinfo{date}{dec}
  \bibinfo{year}{2006}), \bibinfo{pages}{971–987}.
\newblock


\bibitem[\protect\citeauthoryear{Le~Goues, Dewey-Vogt, Forrest, and
  Weimer}{Le~Goues et~al\mbox{.}}{2012}]%
        {legoues:icse:2012}
\bibfield{author}{\bibinfo{person}{Claire Le~Goues}, \bibinfo{person}{Michael
  Dewey-Vogt}, \bibinfo{person}{Stephanie Forrest}, {and}
  \bibinfo{person}{Westley Weimer}.} \bibinfo{year}{2012}\natexlab{}.
\newblock \showarticletitle{A Systematic Study of Automated Program Repair:
  Fixing 55 out of 105 Bugs for \$8 Each}. In
  \bibinfo{booktitle}{\emph{Proceedings of the 34th International Conference on
  Software Engineering}} \emph{(\bibinfo{series}{ICSE})}.
  \bibinfo{publisher}{IEEE Press}, \bibinfo{pages}{3–13}.
\newblock


\bibitem[\protect\citeauthoryear{{Le Goues}, Nguyen, Forrest, and Weimer}{{Le
  Goues} et~al\mbox{.}}{2012}]%
        {legouesNFWTSE2012}
\bibfield{author}{\bibinfo{person}{Claire {Le Goues}}, \bibinfo{person}{ThanhVu
  Nguyen}, \bibinfo{person}{Stephanie Forrest}, {and} \bibinfo{person}{Westley
  Weimer}.} \bibinfo{year}{2012}\natexlab{}.
\newblock \showarticletitle{GenProg: A Generic Method for Automatic Software
  Repair}.
\newblock \bibinfo{journal}{\emph{IEEE Transactions on Software Engineering}}
  \bibinfo{volume}{38} (\bibinfo{year}{2012}), \bibinfo{pages}{54--72}.
\newblock


\bibitem[\protect\citeauthoryear{Li, Liu, Jin, and Umer}{Li
  et~al\mbox{.}}{2020a}]%
        {li:suspicious-return-stmnt:20}
\bibfield{author}{\bibinfo{person}{Guangjie Li}, \bibinfo{person}{Hui Liu},
  \bibinfo{person}{Jiahao Jin}, {and} \bibinfo{person}{Qasim Umer}.}
  \bibinfo{year}{2020}\natexlab{a}.
\newblock \showarticletitle{Deep Learning Based Identification of Suspicious
  Return Statements}. In \bibinfo{booktitle}{\emph{2020 IEEE 27th International
  Conference on Software Analysis, Evolution and Reengineering (SANER)}}.
  \bibinfo{pages}{480--491}.
\newblock


\bibitem[\protect\citeauthoryear{Li, Wang, and Nguyen}{Li
  et~al\mbox{.}}{2020b}]%
        {dlfix}
\bibfield{author}{\bibinfo{person}{Yi Li}, \bibinfo{person}{Shaohua Wang},
  {and} \bibinfo{person}{Tien~N. Nguyen}.} \bibinfo{year}{2020}\natexlab{b}.
\newblock \showarticletitle{DLFix: Context-Based Code Transformation Learning
  for Automated Program Repair}. In \bibinfo{booktitle}{\emph{2020 IEEE/ACM
  42nd International Conference on Software Engineering}}
  \emph{(\bibinfo{series}{ICSE})}. \bibinfo{publisher}{Association for
  Computing Machinery}, \bibinfo{pages}{602–614}.
\newblock


\bibitem[\protect\citeauthoryear{Li, Wang, and Nguyen}{Li
  et~al\mbox{.}}{2022}]%
        {dear-icse-2022}
\bibfield{author}{\bibinfo{person}{Yi Li}, \bibinfo{person}{Shaohua Wang},
  {and} \bibinfo{person}{Tien~N. Nguyen}.} \bibinfo{year}{2022}\natexlab{}.
\newblock \showarticletitle{DEAR: A Novel Deep Learning-Based Approach for
  Automated Program Repair}. In \bibinfo{booktitle}{\emph{Proceedings of the
  44th International Conference on Software Engineering}}
  \emph{(\bibinfo{series}{ICSE '22})}. \bibinfo{publisher}{Association for
  Computing Machinery}, \bibinfo{pages}{511–523}.
\newblock


\bibitem[\protect\citeauthoryear{Li, Zou, Xu, Jin, Zhu, and Chen}{Li
  et~al\mbox{.}}{2021}]%
        {li:sysevr-vulnerability-detection:21}
\bibfield{author}{\bibinfo{person}{Zhen Li}, \bibinfo{person}{Deqing Zou},
  \bibinfo{person}{Shouhuai Xu}, \bibinfo{person}{Hai Jin},
  \bibinfo{person}{Yawei Zhu}, {and} \bibinfo{person}{Zhaoxuan Chen}.}
  \bibinfo{year}{2021}\natexlab{}.
\newblock \showarticletitle{SySeVR: A Framework for Using Deep Learning to
  Detect Software Vulnerabilities}.
\newblock \bibinfo{journal}{\emph{IEEE Transactions on Dependable and Secure
  Computing}} (\bibinfo{year}{2021}), \bibinfo{pages}{1--1}.
\newblock


\bibitem[\protect\citeauthoryear{Li, Zou, Xu, Ou, Jin, Wang, Deng, and
  Zhong}{Li et~al\mbox{.}}{2018}]%
        {VulDeePeckerBinary}
\bibfield{author}{\bibinfo{person}{Zhen Li}, \bibinfo{person}{Deqing Zou},
  \bibinfo{person}{Shouhuai Xu}, \bibinfo{person}{Xinyu Ou},
  \bibinfo{person}{Hai Jin}, \bibinfo{person}{Sujuan Wang},
  \bibinfo{person}{Zhijun Deng}, {and} \bibinfo{person}{Yuyi Zhong}.}
  \bibinfo{year}{2018}\natexlab{}.
\newblock \showarticletitle{VulDeePecker: A Deep Learning-Based System for
  Vulnerability Detection}.
\newblock \bibinfo{journal}{\emph{Proceedings of the Symposium on Network and
  Distributed System Security}} (\bibinfo{year}{2018}).
\newblock


\bibitem[\protect\citeauthoryear{Liu, Koyuncu, Kim, and Bissyand{\'{e}}}{Liu
  et~al\mbox{.}}{2019a}]%
        {liu-avatar-saner-2019}
\bibfield{author}{\bibinfo{person}{Kui Liu}, \bibinfo{person}{Anil Koyuncu},
  \bibinfo{person}{Dongsun Kim}, {and} \bibinfo{person}{Tegawend{\'{e}}~F.
  Bissyand{\'{e}}}.} \bibinfo{year}{2019}\natexlab{a}.
\newblock \showarticletitle{{AVATAR:} Fixing Semantic Bugs with Fix Patterns of
  Static Analysis Violations}. In \bibinfo{booktitle}{\emph{26th {IEEE}
  International Conference on Software Analysis, Evolution and Reengineering}}
  \emph{(\bibinfo{series}{SANER})}. \bibinfo{pages}{456--467}.
\newblock


\bibitem[\protect\citeauthoryear{Liu, Koyuncu, Kim, and Bissyand\'{e}}{Liu
  et~al\mbox{.}}{2019b}]%
        {liu:tbar:issta:2019}
\bibfield{author}{\bibinfo{person}{Kui Liu}, \bibinfo{person}{Anil Koyuncu},
  \bibinfo{person}{Dongsun Kim}, {and} \bibinfo{person}{Tegawend\'{e}~F.
  Bissyand\'{e}}.} \bibinfo{year}{2019}\natexlab{b}.
\newblock \bibinfo{booktitle}{\emph{TBar: Revisiting Template-Based Automated
  Program Repair}}.
\newblock \bibinfo{publisher}{Association for Computing Machinery},
  \bibinfo{pages}{31–42}.
\newblock


\bibitem[\protect\citeauthoryear{Liu and Zhong}{Liu and Zhong}{2018}]%
        {liu-saner-2018}
\bibfield{author}{\bibinfo{person}{Xuliang Liu} {and} \bibinfo{person}{Hao
  Zhong}.} \bibinfo{year}{2018}\natexlab{}.
\newblock \showarticletitle{Mining stackoverflow for program repair}. In
  \bibinfo{booktitle}{\emph{2018 IEEE 25th International Conference on Software
  Analysis, Evolution and Reengineering (SANER)}}. \bibinfo{pages}{118--129}.
\newblock


\bibitem[\protect\citeauthoryear{Lutellier, Pham, Pang, Li, Wei, and
  Tan}{Lutellier et~al\mbox{.}}{2020}]%
        {coconut}
\bibfield{author}{\bibinfo{person}{Thibaud Lutellier},
  \bibinfo{person}{Hung~Viet Pham}, \bibinfo{person}{Lawrence Pang},
  \bibinfo{person}{Yitong Li}, \bibinfo{person}{Moshi Wei}, {and}
  \bibinfo{person}{Lin Tan}.} \bibinfo{year}{2020}\natexlab{}.
\newblock \showarticletitle{CoCoNuT: Combining Context-Aware Neural Translation
  Models Using Ensemble for Program Repair}. In
  \bibinfo{booktitle}{\emph{Proceedings of the 29th ACM SIGSOFT International
  Symposium on Software Testing and Analysis}}
  \emph{(\bibinfo{series}{ISSTA})}. \bibinfo{publisher}{ACM},
  \bibinfo{pages}{101–114}.
\newblock


\bibitem[\protect\citeauthoryear{Mashhadi and Hemmati}{Mashhadi and
  Hemmati}{2021}]%
        {codebert-java-sstubs-21}
\bibfield{author}{\bibinfo{person}{Ehsan Mashhadi} {and} \bibinfo{person}{Hadi
  Hemmati}.} \bibinfo{year}{2021}\natexlab{}.
\newblock \showarticletitle{Applying CodeBERT for Automated Program Repair of
  Java Simple Bugs}. In \bibinfo{booktitle}{\emph{2021 IEEE/ACM 18th
  International Conference on Mining Software Repositories (MSR)}}.
  \bibinfo{pages}{505--509}.
\newblock


\bibitem[\protect\citeauthoryear{Mesbah, Rice, Johnston, Glorioso, and
  Aftandilian}{Mesbah et~al\mbox{.}}{2019}]%
        {deepdelta}
\bibfield{author}{\bibinfo{person}{Ali Mesbah}, \bibinfo{person}{Andrew Rice},
  \bibinfo{person}{Emily Johnston}, \bibinfo{person}{Nick Glorioso}, {and}
  \bibinfo{person}{Edward Aftandilian}.} \bibinfo{year}{2019}\natexlab{}.
\newblock \showarticletitle{DeepDelta: Learning to Repair Compilation Errors}.
  In \bibinfo{booktitle}{\emph{Proceedings of the 2019 27th ACM Joint Meeting
  on European Software Engineering Conference and Symposium on the Foundations
  of Software Engineering}} \emph{(\bibinfo{series}{ESEC/FSE})}.
  \bibinfo{publisher}{Association for Computing Machinery},
  \bibinfo{pages}{925–936}.
\newblock


\bibitem[\protect\citeauthoryear{Monperrus}{Monperrus}{2018}]%
        {monperrus}
\bibfield{author}{\bibinfo{person}{Martin Monperrus}.}
  \bibinfo{year}{2018}\natexlab{}.
\newblock \bibinfo{booktitle}{\emph{The Living Review on Automated Program
  Repair}}.
\newblock \bibinfo{type}{Technical Report}. \bibinfo{institution}{HAL Archives
  Ouvertes}.
\newblock
\urldef\tempurl%
\url{https://hal.archives-ouvertes.fr/hal-01956501}
\showURL{%
\tempurl}


\bibitem[\protect\citeauthoryear{Montemagni and Pirrelli}{Montemagni and
  Pirrelli}{1998}]%
        {contextNLP}
\bibfield{author}{\bibinfo{person}{Simonetta Montemagni} {and}
  \bibinfo{person}{Vito Pirrelli}.} \bibinfo{year}{1998}\natexlab{}.
\newblock \showarticletitle{Augmenting WordNet-like lexical resources with
  distributional evidence. An application-oriented perspective}.
\newblock \bibinfo{journal}{\emph{Proceedings of the COLING/ACL Workshop on Use
  of WordNet in Natural Language Processing Systems}}, \bibinfo{pages}{87--93}.
\newblock


\bibitem[\protect\citeauthoryear{Namavar, Nashid, and Mesbah}{Namavar
  et~al\mbox{.}}{2021}]%
        {controlledAPR}
\bibfield{author}{\bibinfo{person}{Marjane Namavar}, \bibinfo{person}{Noor
  Nashid}, {and} \bibinfo{person}{Ali Mesbah}.}
  \bibinfo{year}{2021}\natexlab{}.
\newblock \bibinfo{title}{A Controlled Experiment of Different Code
  Representations for Learning-Based Bug Repair}.
\newblock
\newblock
\urldef\tempurl%
\url{https://doi.org/10.48550/ARXIV.2110.14081}
\showDOI{\tempurl}


\bibitem[\protect\citeauthoryear{Nguyen, Qi, Roychoudhury, and Chandra}{Nguyen
  et~al\mbox{.}}{2013}]%
        {nguyen:SemFix:icse:2013}
\bibfield{author}{\bibinfo{person}{Hoang Duong~Thien Nguyen},
  \bibinfo{person}{Dawei Qi}, \bibinfo{person}{Abhik Roychoudhury}, {and}
  \bibinfo{person}{Satish Chandra}.} \bibinfo{year}{2013}\natexlab{}.
\newblock \showarticletitle{SemFix: Program Repair via Semantic Analysis}. In
  \bibinfo{booktitle}{\emph{Proceedings of the 2013 International Conference on
  Software Engineering}} \emph{(\bibinfo{series}{ICSE})}.
  \bibinfo{pages}{772–781}.
\newblock


\bibitem[\protect\citeauthoryear{Offutt, Lee, Rothermel, Untch, and
  Zapf}{Offutt et~al\mbox{.}}{1996}]%
        {offutt:wrong-conditions:96}
\bibfield{author}{\bibinfo{person}{A.~Jefferson Offutt}, \bibinfo{person}{Ammei
  Lee}, \bibinfo{person}{Gregg Rothermel}, \bibinfo{person}{Roland~H. Untch},
  {and} \bibinfo{person}{Christian Zapf}.} \bibinfo{year}{1996}\natexlab{}.
\newblock \showarticletitle{An Experimental Determination of Sufficient Mutant
  Operators}.
\newblock \bibinfo{journal}{\emph{ACM Trans. Softw. Eng. Methodol.}}
  \bibinfo{volume}{5}, \bibinfo{number}{2} (\bibinfo{year}{1996}),
  \bibinfo{pages}{99–118}.
\newblock


\bibitem[\protect\citeauthoryear{Pradel and Sen}{Pradel and Sen}{2018}]%
        {pradeloopsla}
\bibfield{author}{\bibinfo{person}{Michael Pradel} {and}
  \bibinfo{person}{Koushik Sen}.} \bibinfo{year}{2018}\natexlab{}.
\newblock \showarticletitle{DeepBugs: A Learning Approach to Name-Based Bug
  Detection}.
\newblock \bibinfo{journal}{\emph{Proceedings of the ACM on Programming
  Languages}} \bibinfo{volume}{2}, \bibinfo{number}{OOPSLA}
  (\bibinfo{year}{2018}), \bibinfo{numpages}{25}~pages.
\newblock


\bibitem[\protect\citeauthoryear{Raffel, Shazeer, Roberts, Lee, Narang, Matena,
  Zhou, Li, and Liu}{Raffel et~al\mbox{.}}{2020}]%
        {t5-model}
\bibfield{author}{\bibinfo{person}{Colin Raffel}, \bibinfo{person}{Noam
  Shazeer}, \bibinfo{person}{Adam Roberts}, \bibinfo{person}{Katherine Lee},
  \bibinfo{person}{Sharan Narang}, \bibinfo{person}{Michael Matena},
  \bibinfo{person}{Yanqi Zhou}, \bibinfo{person}{Wei Li}, {and}
  \bibinfo{person}{Peter~J. Liu}.} \bibinfo{year}{2020}\natexlab{}.
\newblock \showarticletitle{Exploring the Limits of Transfer Learning with a
  Unified Text-to-Text Transformer}.
\newblock \bibinfo{journal}{\emph{Journal of Machine Learning Research}}
  \bibinfo{volume}{21}, \bibinfo{number}{140} (\bibinfo{year}{2020}),
  \bibinfo{pages}{1--67}.
\newblock


\bibitem[\protect\citeauthoryear{Sadowski, Aftandilian, Eagle, Miller-Cushon,
  and Jaspan}{Sadowski et~al\mbox{.}}{2018}]%
        {google:sa}
\bibfield{author}{\bibinfo{person}{Caitlin Sadowski}, \bibinfo{person}{Edward
  Aftandilian}, \bibinfo{person}{Alex Eagle}, \bibinfo{person}{Liam
  Miller-Cushon}, {and} \bibinfo{person}{Ciera Jaspan}.}
  \bibinfo{year}{2018}\natexlab{}.
\newblock \showarticletitle{Lessons from Building Static Analysis Tools at
  Google}.
\newblock \bibinfo{journal}{\emph{Commun. ACM}} \bibinfo{volume}{61},
  \bibinfo{number}{4} (\bibinfo{year}{2018}), \bibinfo{pages}{58--66}.
\newblock


\bibitem[\protect\citeauthoryear{Saha, Lyu, Yoshida, and Prasad}{Saha
  et~al\mbox{.}}{2017}]%
        {saha-elixir-ase-1017}
\bibfield{author}{\bibinfo{person}{Ripon~K. Saha}, \bibinfo{person}{Yingjun
  Lyu}, \bibinfo{person}{Hiroaki Yoshida}, {and} \bibinfo{person}{Mukul~R.
  Prasad}.} \bibinfo{year}{2017}\natexlab{}.
\newblock \showarticletitle{ELIXIR: Effective Object Oriented Program Repair}.
  In \bibinfo{booktitle}{\emph{Proceedings of the 32nd IEEE/ACM International
  Conference on Automated Software Engineering}}
  \emph{(\bibinfo{series}{ASE})}. \bibinfo{pages}{648–659}.
\newblock


\bibitem[\protect\citeauthoryear{Scott, Ranieri, Kot, and Kashyap}{Scott
  et~al\mbox{.}}{2020}]%
        {scott:detecting-swapped-arguments:20}
\bibfield{author}{\bibinfo{person}{Roger Scott}, \bibinfo{person}{Joseph
  Ranieri}, \bibinfo{person}{Lucja Kot}, {and} \bibinfo{person}{Vineeth
  Kashyap}.} \bibinfo{year}{2020}\natexlab{}.
\newblock \showarticletitle{Out of Sight, Out of Place: Detecting and Assessing
  Swapped Arguments}. In \bibinfo{booktitle}{\emph{2020 IEEE 20th International
  Working Conference on Source Code Analysis and Manipulation (SCAM)}}.
  \bibinfo{pages}{227--237}.
\newblock


\bibitem[\protect\citeauthoryear{Son, Shmatikov, and McKinley}{Son
  et~al\mbox{.}}{2013}]%
        {son:access-control-violation:13}
\bibfield{author}{\bibinfo{person}{Sooel Son}, \bibinfo{person}{Vitaly
  Shmatikov}, {and} \bibinfo{person}{Kathryn~S McKinley}.}
  \bibinfo{year}{2013}\natexlab{}.
\newblock \showarticletitle{FixMeUp: Repairing Access-Control Bugs in Web
  Applications}. In \bibinfo{booktitle}{\emph{Network and Distributed System
  Security Symposium (NDSS)} (\bibinfo{edition}{network and distributed system
  security symposium (ndss)} ed.)}.
\newblock


\bibitem[\protect\citeauthoryear{Sotomayor}{Sotomayor}{2022}]%
        {jsScoping}
\bibfield{author}{\bibinfo{person}{Luis Sotomayor}.}
  \bibinfo{year}{2022}\natexlab{}.
\newblock \bibinfo{title}{Avoiding JavaScript Scoping Pitfalls}.
\newblock
  \bibinfo{howpublished}{\url{https://nearsoft.com/blog/avoiding-javascript-scoping-pitfalls}}.
\newblock
\newblock
\shownote{Accessed: 2022-01-06.}


\bibitem[\protect\citeauthoryear{Tan and Roychoudhury}{Tan and
  Roychoudhury}{2015}]%
        {tan:relifix:icse:2015}
\bibfield{author}{\bibinfo{person}{Shin~Hwei Tan} {and} \bibinfo{person}{Abhik
  Roychoudhury}.} \bibinfo{year}{2015}\natexlab{}.
\newblock \showarticletitle{Relifix: Automated Repair of Software Regressions}.
  In \bibinfo{booktitle}{\emph{Proceedings of the 37th International Conference
  on Software Engineering - Volume 1}} \emph{(\bibinfo{series}{ICSE})}.
  \bibinfo{pages}{471–482}.
\newblock


\bibitem[\protect\citeauthoryear{Tufano, Watson, Bavota, Penta, White, and
  Poshyvanyk}{Tufano et~al\mbox{.}}{2019}]%
        {tufano:tosem:19}
\bibfield{author}{\bibinfo{person}{Michele Tufano}, \bibinfo{person}{Cody
  Watson}, \bibinfo{person}{Gabriele Bavota}, \bibinfo{person}{Massimiliano~Di
  Penta}, \bibinfo{person}{Martin White}, {and} \bibinfo{person}{Denys
  Poshyvanyk}.} \bibinfo{year}{2019}\natexlab{}.
\newblock \showarticletitle{An Empirical Study on Learning Bug-Fixing Patches
  in the Wild via Neural Machine Translation}.
\newblock \bibinfo{journal}{\emph{ACM Transactions on Software Engineering and
  Methodology}} (\bibinfo{year}{2019}), \bibinfo{numpages}{29}~pages.
\newblock


\bibitem[\protect\citeauthoryear{Weimer, Fry, and Forrest}{Weimer
  et~al\mbox{.}}{2013}]%
        {weimer:ase:2013}
\bibfield{author}{\bibinfo{person}{Westley Weimer}, \bibinfo{person}{Zachary~P.
  Fry}, {and} \bibinfo{person}{Stephanie Forrest}.}
  \bibinfo{year}{2013}\natexlab{}.
\newblock \showarticletitle{Leveraging Program Equivalence for Adaptive Program
  Repair: Models and First Results}. In \bibinfo{booktitle}{\emph{Proceedings
  of the 28th IEEE/ACM International Conference on Automated Software
  Engineering}} \emph{(\bibinfo{series}{ASE})}. \bibinfo{pages}{356–366}.
\newblock


\bibitem[\protect\citeauthoryear{Weimer, Nguyen, Le~Goues, and Forrest}{Weimer
  et~al\mbox{.}}{2009}]%
        {weimer:icse:2009}
\bibfield{author}{\bibinfo{person}{Westley Weimer}, \bibinfo{person}{ThanhVu
  Nguyen}, \bibinfo{person}{Claire Le~Goues}, {and} \bibinfo{person}{Stephanie
  Forrest}.} \bibinfo{year}{2009}\natexlab{}.
\newblock \showarticletitle{Automatically Finding Patches Using Genetic
  Programming}. In \bibinfo{booktitle}{\emph{Proceedings of the 31st
  International Conference on Software Engineering}}
  \emph{(\bibinfo{series}{ICSE})}. \bibinfo{publisher}{IEEE Computer Society},
  \bibinfo{pages}{364–374}.
\newblock


\bibitem[\protect\citeauthoryear{Weiser}{Weiser}{1979}]%
        {weiser}
\bibfield{author}{\bibinfo{person}{Mark~D. Weiser}.}
  \bibinfo{year}{1979}\natexlab{}.
\newblock \showarticletitle{Program slices: formal, psychological, and
  practical investigations of an automatic program abstraction method}. In
  \bibinfo{booktitle}{\emph{Proceedings of the 5th International Conference on
  Software Engineering}} \emph{(\bibinfo{series}{ICSE})}.
  \bibinfo{pages}{439–449}.
\newblock


\bibitem[\protect\citeauthoryear{Wen, Chen, Wu, Hao, and Cheung}{Wen
  et~al\mbox{.}}{2018}]%
        {capgen}
\bibfield{author}{\bibinfo{person}{Ming Wen}, \bibinfo{person}{Junjie Chen},
  \bibinfo{person}{Rongxin Wu}, \bibinfo{person}{Dan Hao}, {and}
  \bibinfo{person}{Shing-Chi Cheung}.} \bibinfo{year}{2018}\natexlab{}.
\newblock \showarticletitle{Context-Aware Patch Generation for Better Automated
  Program Repair}. In \bibinfo{booktitle}{\emph{Proceedings of the 40th
  International Conference on Software Engineering}}
  \emph{(\bibinfo{series}{ICSE})}. \bibinfo{pages}{1--11}.
\newblock


\bibitem[\protect\citeauthoryear{Xiao, Ahmed, Song, Ge, Viswanath, and
  Yao}{Xiao et~al\mbox{.}}{2021}]%
        {slicing:apirecommendation:2021}
\bibfield{author}{\bibinfo{person}{Ya Xiao}, \bibinfo{person}{Salman Ahmed},
  \bibinfo{person}{Wenjia Song}, \bibinfo{person}{Xinyang Ge},
  \bibinfo{person}{Bimal Viswanath}, {and} \bibinfo{person}{Danfeng Yao}.}
  \bibinfo{year}{2021}\natexlab{}.
\newblock \showarticletitle{Embedding Code Contexts for Cryptographic API
  Suggestion: New Methodologies and Comparisons}.
\newblock \bibinfo{journal}{\emph{arXiv preprint arXiv:2103.08747}}
  (\bibinfo{year}{2021}).
\newblock


\bibitem[\protect\citeauthoryear{Xuan, Martinez, DeMarco, Clement, Marcote,
  Durieux, Le~Berre, and Monperrus}{Xuan et~al\mbox{.}}{2017}]%
        {xuan:wrong-conditions:17}
\bibfield{author}{\bibinfo{person}{Jifeng Xuan}, \bibinfo{person}{Matias
  Martinez}, \bibinfo{person}{Favio DeMarco}, \bibinfo{person}{Maxime Clement},
  \bibinfo{person}{Sebastian~Lamelas Marcote}, \bibinfo{person}{Thomas
  Durieux}, \bibinfo{person}{Daniel Le~Berre}, {and} \bibinfo{person}{Martin
  Monperrus}.} \bibinfo{year}{2017}\natexlab{}.
\newblock \showarticletitle{Nopol: Automatic Repair of Conditional Statement
  Bugs in Java Programs}.
\newblock \bibinfo{journal}{\emph{IEEE Trans. Softw. Eng.}}
  \bibinfo{volume}{43}, \bibinfo{number}{1} (\bibinfo{year}{2017}),
  \bibinfo{pages}{34–55}.
\newblock


\bibitem[\protect\citeauthoryear{Ye, Martinez, and Monperrus}{Ye
  et~al\mbox{.}}{2022}]%
        {rewardrepair-icse-2022}
\bibfield{author}{\bibinfo{person}{He Ye}, \bibinfo{person}{Matias Martinez},
  {and} \bibinfo{person}{Martin Monperrus}.} \bibinfo{year}{2022}\natexlab{}.
\newblock \showarticletitle{Neural Program Repair with Execution-Based
  Backpropagation}. In \bibinfo{booktitle}{\emph{Proceedings of the 44th
  International Conference on Software Engineering}}
  \emph{(\bibinfo{series}{ICSE '22})}. \bibinfo{publisher}{Association for
  Computing Machinery}, \bibinfo{pages}{1506–1518}.
\newblock


\bibitem[\protect\citeauthoryear{Yu, Alkhalaf, and Bultan}{Yu
  et~al\mbox{.}}{2011}]%
        {yu:string-sanitization:11}
\bibfield{author}{\bibinfo{person}{Fang Yu}, \bibinfo{person}{Muath Alkhalaf},
  {and} \bibinfo{person}{Tevfik Bultan}.} \bibinfo{year}{2011}\natexlab{}.
\newblock \showarticletitle{Patching Vulnerabilities with Sanitization
  Synthesis}. In \bibinfo{booktitle}{\emph{Proceedings of the 33rd
  International Conference on Software Engineering}}
  \emph{(\bibinfo{series}{ICSE '11})}. \bibinfo{publisher}{Association for
  Computing Machinery}, \bibinfo{pages}{251–260}.
\newblock


\bibitem[\protect\citeauthoryear{Zhang, Wang, Zhang, Li, and Jin}{Zhang
  et~al\mbox{.}}{2022}]%
        {het-graph-icpc-2022}
\bibfield{author}{\bibinfo{person}{Kechi Zhang}, \bibinfo{person}{Wenhan Wang},
  \bibinfo{person}{Huangzhao Zhang}, \bibinfo{person}{Ge Li}, {and}
  \bibinfo{person}{Zhi Jin}.} \bibinfo{year}{2022}\natexlab{}.
\newblock \showarticletitle{Learning to Represent Programs with Heterogeneous
  Graphs}. In \bibinfo{booktitle}{\emph{2022 IEEE/ACM 30th International
  Conference on Program Comprehension (ICPC)}}. \bibinfo{pages}{378--389}.
\newblock


\bibitem[\protect\citeauthoryear{Zhou, Cui, Hu, Zhang, Yang, Liu, Wang, Li, and
  Sun}{Zhou et~al\mbox{.}}{2020}]%
        {gnn}
\bibfield{author}{\bibinfo{person}{Jie Zhou}, \bibinfo{person}{Ganqu Cui},
  \bibinfo{person}{Shengding Hu}, \bibinfo{person}{Zhengyan Zhang},
  \bibinfo{person}{Cheng Yang}, \bibinfo{person}{Zhiyuan Liu},
  \bibinfo{person}{Lifeng Wang}, \bibinfo{person}{Changcheng Li}, {and}
  \bibinfo{person}{Maosong Sun}.} \bibinfo{year}{2020}\natexlab{}.
\newblock \showarticletitle{Graph neural networks: A review of methods and
  applications}.
\newblock \bibinfo{journal}{\emph{AI Open}} (\bibinfo{year}{2020}),
  \bibinfo{pages}{57--81}.
\newblock


\bibitem[\protect\citeauthoryear{Zhou, Liu, Siow, Du, and Liu}{Zhou
  et~al\mbox{.}}{2019}]%
        {devign-nips-vulnerability-detection-gnn}
\bibfield{author}{\bibinfo{person}{Yaqin Zhou}, \bibinfo{person}{Shangqing
  Liu}, \bibinfo{person}{Jingkai Siow}, \bibinfo{person}{Xiaoning Du}, {and}
  \bibinfo{person}{Yang Liu}.} \bibinfo{year}{2019}\natexlab{}.
\newblock \bibinfo{booktitle}{\emph{Devign: Effective Vulnerability
  Identification by Learning Comprehensive Program Semantics via Graph Neural
  Networks}}.
\newblock \bibinfo{publisher}{Curran Associates Inc.}
\newblock


\bibitem[\protect\citeauthoryear{Zhu, Sun, Xiao, Zhang, Yuan, Xiong, and
  Zhang}{Zhu et~al\mbox{.}}{2021}]%
        {recoder-fse-2021}
\bibfield{author}{\bibinfo{person}{Qihao Zhu}, \bibinfo{person}{Zeyu Sun},
  \bibinfo{person}{Yuan-an Xiao}, \bibinfo{person}{Wenjie Zhang},
  \bibinfo{person}{Kang Yuan}, \bibinfo{person}{Yingfei Xiong}, {and}
  \bibinfo{person}{Lu Zhang}.} \bibinfo{year}{2021}\natexlab{}.
\newblock \showarticletitle{A Syntax-Guided Edit Decoder for Neural Program
  Repair}. In \bibinfo{booktitle}{\emph{Proceedings of the 29th ACM Joint
  Meeting on European Software Engineering Conference and Symposium on the
  Foundations of Software Engineering}} \emph{(\bibinfo{series}{ESEC/FSE
  2021})}. \bibinfo{publisher}{Association for Computing Machinery},
  \bibinfo{pages}{341–353}.
\newblock


\bibitem[\protect\citeauthoryear{Zou, Wang, Xu, Li, and Jin}{Zou
  et~al\mbox{.}}{2019}]%
        {VulDeePeckerMulti}
\bibfield{author}{\bibinfo{person}{Deqing Zou}, \bibinfo{person}{Sujuan Wang},
  \bibinfo{person}{Shouhuai Xu}, \bibinfo{person}{Zhen Li}, {and}
  \bibinfo{person}{Hai Jin}.} \bibinfo{year}{2019}\natexlab{}.
\newblock \showarticletitle{{$\mu$}VulDeePecker: A Deep Learning-Based System
  for Multiclass Vulnerability Detection}.
\newblock \bibinfo{journal}{\emph{IEEE Transactions on Dependable and Secure
  Computing}} (\bibinfo{year}{2019}), \bibinfo{pages}{1–1}.
\newblock


\end{thebibliography}

\end{document}